\documentstyle[prd,aps,floats,psfig,amsfonts]{revtex}
\input epsf.sty

\newcommand{\n}{{\mathbf{n}}}
\newcommand{\U}{\underline}
\newcommand{\f}{\frac}

\newcommand{\bb}{\bibitem}
\newcommand{\BF}{\begin{figure}}
\newcommand{\EF}{\end{figure}}
\newcommand{\BE}{\begin{equation}}
\newcommand{\EE}{\end{equation}}
\newcommand{\BEA}{\begin{eqnarray}}
\newcommand{\EEA}{\end{eqnarray}}
\tighten
\begin{document}
\draft
\title{Geometric Gaussianity and Non-Gaussianity in the Cosmic
Microwave Background}
\author{Kaiki Taro Inoue}
\address{Yukawa Institute for Theoretical Physics, Kyoto University,
Kyoto 606-8502, Japan}
\date{\today}
\maketitle
\begin{abstract}
In this paper, Gaussianity of eigenmodes and 
non-Gaussianity in the Cosmic Microwave Background (CMB)
temperature fluctuations in
two smallest compact hyperbolic (CH) models are investigated. 
First, it is numerically found that the expansion coefficients of 
low-lying eigenmodes on the two CH manifolds behave as if 
they are Gaussian random numbers 
at almost all the places.
Next, non-Gaussianity of the temperature fluctuations in the 
($l,m$) space in these models
is studied. Assuming that the initial fluctuations are 
Gaussian, the real expansion coefficients $b_{l m}$ of the temperature 
fluctuations in the sky are found to be 
distinctively non-Gaussian. In particular, 
the cosmic variances are found to be much larger than that
for Gaussian models. On the other hand, the 
anisotropic structure is vastly erased if one averages the 
fluctuations at a number of different observing points because of the 
Gaussian pseudo-randomness of the eigenmodes. Thus the dominant
contribution to the two-point correlation functions comes from the
isotropic terms described by the angular power spectra $C_l$. 
Finally, topological quantities: the total length and the genus 
of isotemperature contours are
investigated. The variances of total length and genus
at high and low threshold levels 
are found to be considerably larger than that of
Gaussian models while the means almost agree with them.
\end{abstract}

\pacs{PACS Numbers : 98.70.Vc, 98.80.Hw}
\begin{picture}(0,0)
\put(410,280)
{YITP-00-11}
\end{picture}
\section{introduction}
\indent
In recent years, locally isortopic and homogeneous 
Friedmann-Robertson-Walker (FRW) models with non-trivial
topology have attracted much attention. In the standard scenario, 
simply-connectivity of the spatial hypersurface is assumed for
simplicity.  However, the Einstein equations, being local equations, 
do not fix the global topology of the spacetime. 
In other words, a wide variety of 
topologically distinct spacetimes with the same local geometry 
described by a local metric element remain unspecified (see
\cite{Lachieze} for review on the cosmological topology). 
The determination of the global
topology of the universe is one of the most important problem of the
modern observational cosmology. 
\\
\indent
For flat models without the cosmological constant,
severest constraints have been obtained by using the COBE
DMR data. The suppression of the fluctuations on scales beyond the 
topological identification scale $L$ leads to the decrease of the 
angular power spectra $C_l$ of the Cosmic Microwave Background (CMB)
temperature fluctuations on large angular scales 
which puts a lower bound $L\!\ge\! 2400~ h^{-1}
$Mpc (with $h\!=\!H_{0}/100~\mathrm{kms^{-1}Mpc^{-1}
}$) for a compact flat 3-torus model without the cosmological constant
\cite{Stevens,Oliveira}. Similar constraints have
been obtained for other compact flat models \cite{Levin}. 
The maximum expected number
of copies of the fundamental domain (cell) inside the last scattering
surface is approximately 8 for the 3-torus model. 
\\
\indent
In contrast, for low density models, the constraint could be
considerably milder than the locally isotropic and homogeneous 
flat (Einstein-de-Sitter) models
since a bulk of large-angle CMB fluctuations 
can be produced by the so-called (late) integrated Sachs-Wolfe effect
(ISW) \cite{HSS,Cornish2} which is the gravitational blueshift
effect of the free streaming photons by the decay of the gravitational 
potential. As the gravitational potential decays in either 
$\Lambda$-dominant epoch or curvature dominant epoch, 
the free streaming photons  
with large wavelength (the light travel time across the wavelength 
is greater than or comparable to
the decay time) that climbed a potential well at the last scattering 
experience blueshifts due to the contraction of the comoving space
along the trajectories of the photons.
Because the angular sizes of the fluctuations produced at late time 
are large, the suppression of the fluctuations on scale larger than the 
topological identification scale does not lead to a
significant suppression of the large-angle power if the 
ISW effect is dominant.
Recent works \cite{Aurich3,Inoue2,Inoue3,Cornish1} have shown that 
the large-angle power ($2\!\le\!l\!\le\!20$) are completely consistent with
the COBE DMR data for compact hyperbolic (CH) models which 
include a small CH orbifold, the Weeks and the Thurston manifolds 
with volume $0.72, 0.94$ and $0.98$ in unit of the cube of the
curvature radius, respectively. Note that the Weeks manifold is the
smallest and the Thurston manifolds is the second smallest 
in the known CH manifolds.
For instance, the number
of copies of the fundamental domain inside the last scattering
surface at present is approximately 190 for a Weeks model 
with $\Omega_{0}\!=\!0.3$.
\\
\indent
If the space is negatively curved, for a fixed number of
the copies of the fundamental domain inside the present horizon, 
the large-angle fluctuations can be produced much effectively. 
In negatively curved spaces (hyperbolic spaces), trajectories of 
photons subtend a much
smaller angle in the sky for a given scale. In other words, 
for a given angle of a pair
of two photon trajectories, the physical distance of the 
trajectories is much greater than
that in a flat space. Therefore, even if there is a number of copies
of the fundamental domain which intersect the last scattering surface, 
the number of copies which intersect the wave front (a sphere with
$z\!=\!\textrm{const.}$) 
of the free streaming photons is
exponentially decreased at late time when the large-angle
fluctuations are produced due to the ISW effect.
\\
\indent
However, one may not be satisfied with the constraints using only the 
angular power spectrum $C_{l}$ since it contains only isotropic
information of the ensemble averaged temperature fluctuations \cite{Bond}. 
If they have anisotropic 
structures, non-Gaussian signatures must be revealed.
In fact, the global isotropy 
of the locally isotropic and homogeneous 
FRW models is generally
broken. For instance, a flat 3-torus obtained by identifying
the opposite faces of a cube is obviously anisotropic 
at any points. Thus the temperature
fluctuations averaged over the initial conditions 
in these multiply-connected FRW models are no 
longer $SO(3)$ invariant at a certain point.
The temperature
fluctuations on the sky are written in terms of (real) 
spherical harmonics $Q_{lm}(\n)$ as 
\BE
\f{\Delta T}{T}(\n)=\sum_{l}~\sum_{m=-l}^{l} b_{lm} Q_{lm}(\n).
\EE
If the distribution functions of the real 
expansion coefficients $b_{lm}$ are $SO(3)$ invariant, 
the temperature fluctuations must be Gaussian provided that 
$b_{lm}$'s are independent random numbers \cite{Mag1}. Therefore, the
temperature fluctuations at a certain point in the multiply-connected 
FRW models are not Gaussian if $b_{lm}$'s are independent. 
\\
\indent
For the simplest flat 3-torus models 
(without rotations in the identification maps) which are 
globally homogeneous, it is sufficient to choose one observing 
point and estimate how the power is distributed among the $m$'s for 
a given angular scale $l$ in order to see the effect of the global
anisotropy.
However, in general, one must consider an ensemble of fluctuations
at different observing points because of the spatial (global)
inhomogeneity. Previous analyses have not fully investigated the 
dependence of the temperature fluctuations on choice of the 
observing points.
\\
\indent
Lack of analytical results on the eigenmodes 
makes it difficult to investigate the nature of the temperature fluctuations
in CH  models.
However, we may expect a high degree of complexity in the 
eigenmodes since the corresponding
classical systems (geodesic flows) are strongly chaotic. 
In fact, it has been numerically found that
the expansion coefficients of the low-lying eigenmodes on the 
Thurston manifold at the point where the injectivity radius 
is maximal are Gaussian 
pseudo-random numbers \cite{Inoue1} 
which supports the previous analysis of the excited states (higher 
modes) of a two-dimensional
asymmetrical CH model \cite{Aurich2}. We have put a prefix ''pseudo'' since 
the eigenmodes are actually constrained by the periodic boundary conditions. 
These results imply that the statistical properties of the eigenmodes
on CH spaces (orbifolds and manifolds) can be described 
by random-matrix theory (RMT)\cite{Meh,Boh}. 
An investigation of the dependence of the property on the observing
points is also important since CH spaces have symmetries (isometric
groups) which may veil the random feature of the eigenmodes. In this
paper, a detailed analysis on the statistical property of low-lying 
eigenmodes on the Weeks and the Thurston manifolds is conducted.
\\
\indent
Assuming that the eigenmodes are Gaussian, one can expect that 
the anisotropic structure in the $(l,m)$ space is 
vastly erased when one averages the
fluctuations over the space.
This seems to be a paradox since the CH spaces are actually globally
anisotropic. However, one should consider a spatial average of 
fluctuations with different initial conditions if
one believes the Copernican principle that we are not in the center
of the universe.
Even if the space is anisotropic 
at a certain point, the averaged fluctuations may 
look isotropic by considering
an ensemble of fluctuations at all the possible observing points. 
Note that the eigenmodes on CH spaces
have no particular directions if they are Gaussian.
\\
\indent
If the initial fluctuations are constant for each eigenmode, 
as we shall see, the Gaussian 
randomness of the temperature fluctuations can be solely 
attributed to the Gaussian pseudo-randomness of the eigenmodes.
In this case, the Gaussian randomness of the temperature fluctuations has its
origin in the geometrical property of the space
(\textit{Geometric Gaussianity}). Choosing an observing 
point is equivalent to fixing a certain initial condition.
However, it is much natural to assume that the initial fluctuations are
also random Gaussian as the standard inflationary scenarios predict.
Then the temperature fluctuations may not obey the Gaussian statistics
because they are written in terms of products of two different
independent Gaussian numbers rather than sums while they remain almost 
spatially isotropic if averaged over the space.
\\
\indent
In this paper, Gaussianity of eigenmodes and non-Gaussianity in
the CMB for two smallest CH models (the Weeks and the Thurston models) 
are investigated. In Sec.~II, numerical results on Gaussianity of eigenmodes 
are shown and we discuss to what extent the results are generic.
In Sec.~III, we study the non-Gaussian behavior of the
temperature fluctuations in the ($l,m$) space. In Sec.~IV,
topological quantities (total length and genus) of isotemperature 
contours are numerically simulated for studying the non-Gaussian behavior
in the real space. Finally, we summarize our conclusions in Sec.~V.
\section{GEOMETRIC GAUSSIANITY}
In locally isotropic and homogeneous FRW background spaces, 
each type (scalar, vector and tensor) of
first-order perturbations can be decomposed into a decoupled set of
equations. In order to solve the decomposed linearly perturbed
Einstein equations, it is useful to expand the  
perturbations in terms of eigenmodes
of the Laplacian which satisfies the Helmholtz equation with certain
boundary conditions,
\BE
(\nabla^2+k^2)u_{k}(x)=0,
\EE
since each eigenmode evolves independently in the linear approximation.
Then one can easily see that the time evolution of the 
perturbations in the multiply-connected locally isotropic and
homogeneous FRW
spaces coincide with that in the FRW spaces while 
the global structure of the background space is described solely by these 
eigenmodes. 
\\
\indent
Unfortunately, no analytical expressions of eigenmodes on CH spaces
have been known. Nevertheless, the correspondence
between classical and quantum mechanics may provide us a clue for
understanding the generic property of the eigenmodes. If one
recognizes the Laplacian as the Hamiltonian in a quantum system, 
each eigenmode can be interpreted as a wavefunction in a stationary state.   
Because classical dynamical systems (=geodesic 
flows) on CH spaces are strongly chaotic (or more precisely they are K-systems 
with ergodicity, mixing and Bernoulli properties \cite{Balazs}), one can expect
a high degree of complexity for each eigenstate. Imprint of the chaos 
in the classical systems may be hidden in the 
quantum counterparts.
In fact, in many cases, the short-range correlations observed
in the eigenvalues (energy states) have been found to be consistent 
with the universal prediction of RMT for three universality 
classes:the Gaussian 
orthogonal ensemble(GOE), the Gaussian unitary ensemble(GUE) and the
Gaussian symplectic ensemble (GSE)\cite{Meh,Boh}. 
In our case the statistical properties are described by GOE (which
consist of real symmetric $N \!\times\!N$ matrices $H$ which obey the
Gaussian distribution $\propto \exp{(-\textrm{Tr}H^2/(4a^2))}$ 
(where $a$ is a constant) as the systems possess a time-reversal
symmetry. RMT also predicts that
the squared expansion coefficients of an eigenstate with respect to a 
generic basis are distributed as Gaussian random numbers \cite{Bro}. 
Unfortunately, no analytic forms of generic bases(=eigenmodes) are
known for CH spaces which seems to be an intractable problem.
However, if the eigenmodes are continued onto the universal covering
space by the periodic boundary conditions, they can be written in terms
of a ''generic'' basis on the universal covering space (=3-hyperboloid
$H^3$). In pseudospherical coordinates ($R,\chi,\theta,\phi$), 
the eigenmodes are written in terms of complex expansion 
coefficients $\xi_{\nu l m}$
and eigenmodes on the universal covering space, 
\BE
u_\nu=\sum_{l m} \xi_{\nu l m}\,X_{\nu l}(\chi) Y_{l m}(\theta,\phi),
\label{eq:u}
\EE
where $\nu=\sqrt{k^2-1}$, $X_{\nu l}$ and $ Y_{l m}$ denote the
radial eigenfunction and (complex) spherical harmonic on the pseudosphere 
with radius $R$, respectively. Then the real expansion 
coefficients $a_{\nu l m}$ are given by
\begin{eqnarray}
\nonumber
a_{\nu 0  0}&=&-\textrm{Im}(\xi_{\nu 0  0}),~~~~
a_{\nu l  0}=\sqrt{c_{\nu l}}\textrm{Re}(\xi_{\nu l 0}),
\\
\nonumber
a_{\nu l  m}&=&\sqrt{2}\textrm{Re}(\xi_{\nu l m}),~~m>0,
\\
a_{\nu l  m}&=&-\sqrt{2}\textrm{Im}(\xi_{\nu l -m}),~~m<0,
\end{eqnarray}
where
\begin{eqnarray}
\nonumber
c_{\nu l}&=&\f{2}{(1+\textrm{Re}(F(\nu,l)))},
\\
F(\nu, l)&=&\f{\Gamma(l+\nu i+1)}{\Gamma(\nu i)}
\f{\Gamma(-\nu i)}{\Gamma(l-\nu i+1)}.
\end{eqnarray}
\\
\indent
In this paper, the low-lying 
eigenmodes ($k<13$) on the Weeks and Thurston
manifolds are numerically computed by the direct boundary 
element method.  The identification matrices of the Dirichlet domains 
are obtained by a computer program ``SnapPea'' by J. R. Weeks \cite{Weeks1}. 
The computed eigenvalues are 
well consistent with that in the previous literature 
\cite{Inoue1,Cornish3}.  
The estimated errors in $k$ are within $0.01$. However, the
last digits in $k$ may be incorrect.  
$a_{\nu l m}$ 's can be promptly obtained after the normalization and
orthogonalization of these eigenmodes. The orthogonalization is
achieved at the level of $10^{-3}$ to $10^{-4}$ (for the inner product
of the normalized eigenmodes) which implies that each eigenmode is 
computed with relatively high accuracy.
\BF
\centerline{\psfig{figure=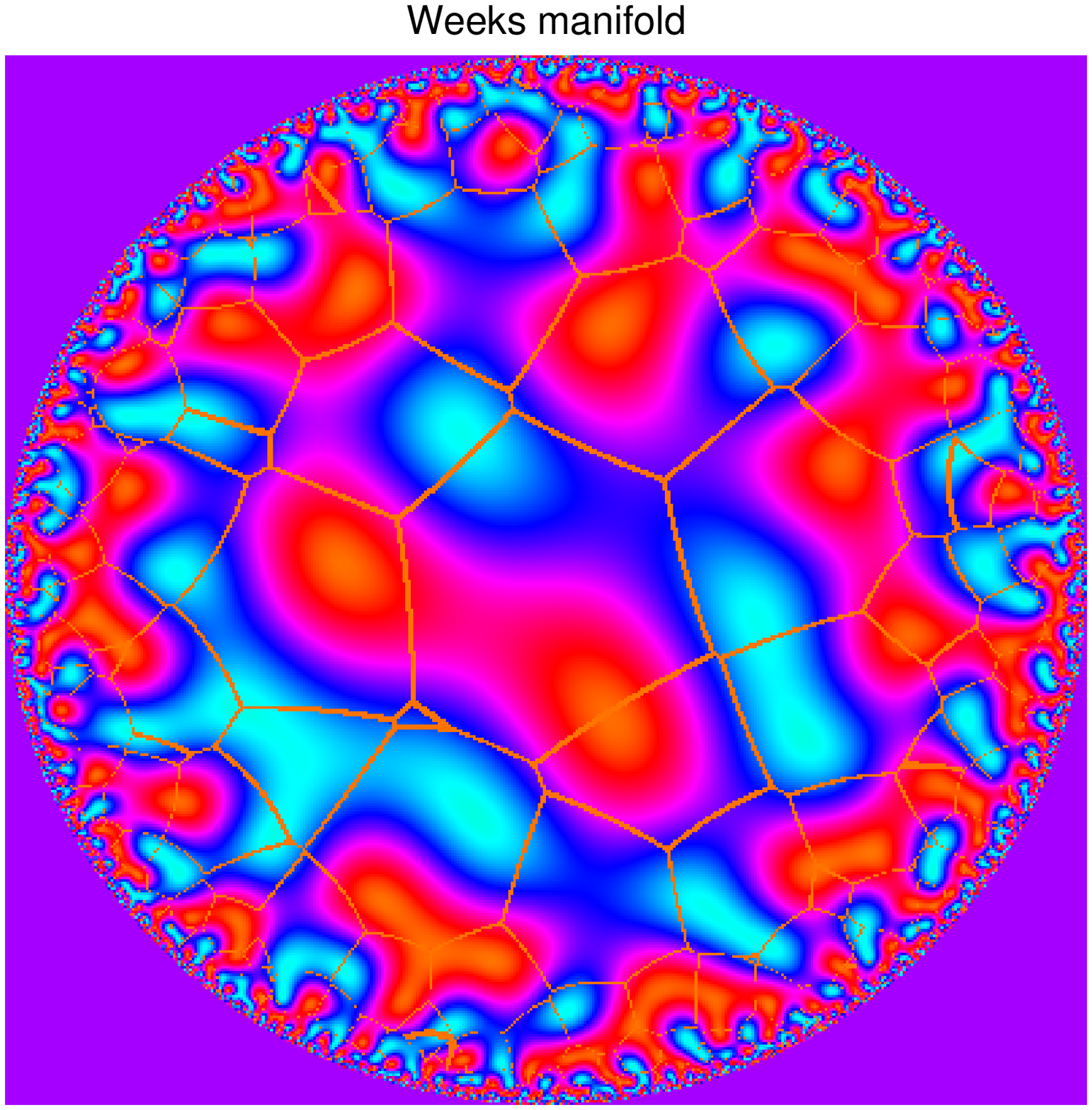,width=17cm}}
\caption{The lowest eigenmode $k\!=\!5.268$ on the Weeks manifold 
continued onto the Poincar$\acute{\textrm{e}}$ 
ball and the boundaries of the copied Dirichlet domains (\textit{solid
curves}) plotted on a slice $z\!=\!0$.}
\label{fig:EMW}
\EF
\BF
\centerline{\psfig{figure=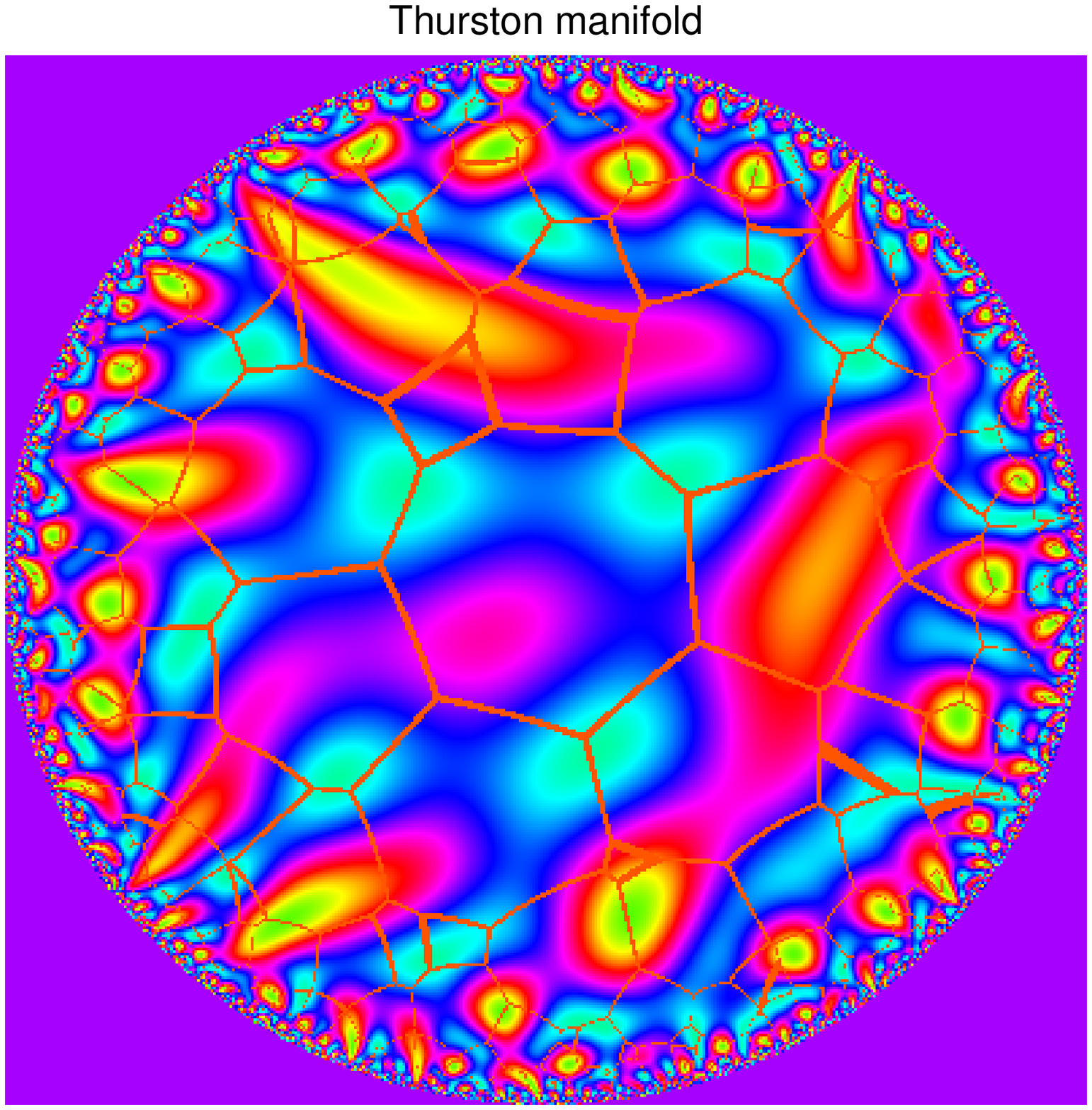,width=17cm}}
\caption{The lowest eigenmode $k\!=\!5.404$ on the Thurston manifold 
continued onto the Poincar$\acute{\textrm{e}}$ 
ball and the boundaries of the copied Dirichlet domains (\textit{solid
curves}) plotted on a slice $z\!=\!0$.}
\label{fig:EMT}
\EF
In Fig.1 and Fig.2, one can see a high degree of complexity in the lowest 
eigenmodes on the Poincar$\acute{\textrm{e}}$ ball which is isometric
to the universal covering space $H^3$ whose coordinates are
given by
\BE
x=R \tanh \f{\chi}{2} \sin \theta \cos \phi, ~~
y=R \tanh \f{\chi}{2} \sin \theta \sin \phi, ~~
z=R \tanh \f{\chi}{2} \cos \theta.
\EE
Replacing $\tanh \f{\chi}{2}$ by $\tanh \chi$ for each
coordinate, one obtains the Klein (projective) coordinates.
In the Poincar$\acute{\textrm{e}}$ coordinates, angles of 
geodesics coincide with that of Euclidean ones. 
In the Klein coordinates, all geodesics are straight 
lines while angles does not coincide with that of Euclidean ones.
\BF[tpb]
\centerline{\psfig{figure=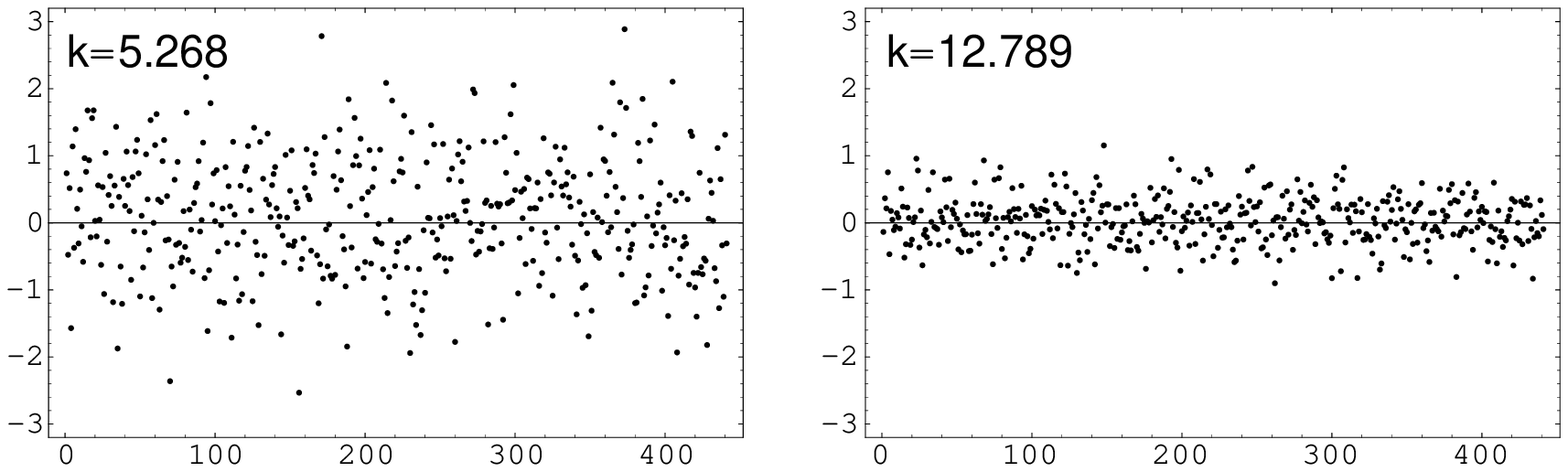,width=17cm}}
\caption{Plots of $a_{\nu l m}$'s which are ordered 
as $l(l+1)+m+1, 0\le l \le 20$ for eigenmodes 
$k\!=\!5.268$(left) and $k\!=\!12.789$(right) on the Weeks manifold
at a point which is randomly chosen.}
\label{fig:anulmW}
\EF
In what follows $R$ is normalized to 1 without loss of generality.
\\
\indent
In Fig.3, one can see that the distribution of 
$a_{\nu l m}$'s which are ordered as $l(l+1)+m+1$ 
are qualitatively random.
In order to estimate the randomness 
quantitatively, we consider a cumulative distribution of 
\begin{equation}
b_{\nu l m}=\f{|a_{\nu l m}-\bar{a}_{\nu}|^2}{\sigma_{\nu}^2}
\end{equation}
where $\bar{a}_{\nu}$ is the mean of $a_{\nu
l m}$'s and $\sigma_{\nu}^2$ is the variance.
If $a_{\nu l m}$'s are Gaussian then $b_{\nu l m}$ 's obey
a $\chi^2$ distribution $P(x)=(1/2)^{1/2}\Gamma(1/2)x^{-1/2}e^{-x/2}$
with 1 degree of freedom. To test the goodness of fit between the
the theoretical cumulative distribution $I(x)$ and the empirical 
cumulative distribution function $I_N(x)$, we use 
the Kolmogorov-Smirnov statistic $D_N$ which is the least upper bound
of all pointwise differences $|I_N(x)-I(x)|$
\cite{Hog},
\begin{equation}
D_N\equiv \sup_{x} |I_N(x)-I(x)|.
\end{equation}
$I_N(x)$ is defined as
\begin{eqnarray}
 I_N(x)&=&\left\{ \begin{array}{@{\,}ll}
0, & x<y_1,
\\
j/N,~~& y_j \leq x < y_{j+1},~~~~j=1,2,\ldots,N\!-\!1,
\\
1, & y_N \leq x, 
\end{array}
\right. 
\end{eqnarray}
where $y_1<y_2< \ldots <y_N$ are the computed values of a random
sample which consists of $N$ elements. 
For random variables $D_N$ for any $z>0$, it can be shown that 
the probability of $D_N\!<\!d$ is given by \cite{Bir}
\begin{equation}
\lim_{N \rightarrow \infty} ~P(D_N<d=z N^{-1/2})=L(z),\label{eq:P} 
\end{equation}  
where
\begin{equation}
L(z)=1-2 \sum_{j=1}^{\infty} (-1)^{j-1} e^{-2j^2 z^2}.
\end{equation}
From the observed maximum difference $D_N\!=\!d$, we obtain 
the significance level $\alpha_D\!=\!1-P$ which is equal to the probability 
of $D_N\!>\!d$. If $\alpha_D$ is found to be large enough, 
the hypothesis $I_N(x)\!\neq\! I(x)$ is not verified. 
The significance levels $\alpha_N$ for $0 \!\leq\! l \!\leq\! 20$ for
eigenmodes $k\!<\!13$ on the Thurston manifold are shown in table 1. 
\begin{table}
\begin{center}
\begin{tabular}{cccc}   
\multicolumn{1}{c}{k} &
\multicolumn{1}{c}{$\alpha_D$} &
\multicolumn{1}{c}{k} &
\multicolumn{1}{c}{$\alpha_D$} 
\\ \hline
5.404&0.98          &10.686b& $7.9\times 10^{-4}$ \\ \hline
5.783&0.68          &10.737 & 0.96 \\ \hline
6.807a &0.52       &10.830 & 0.67 \\ \hline
6.807b & $7.1\times 10^{-4}$  &11.103a & 0.041 \\ \hline
6.880&1.00          &11.103b & $8.8\times 10^{-15}$ \\ \hline
7.118&0.79          &11.402 & 0.98 \\ \hline
7.686a&0.26         &11.710 & 0.92 \\ \hline
7.686b& $2.3\times 10^{-8}$ &11.728 & 0.93 \\ \hline
8.294&0.45          &11.824 & 0.31 \\ \hline
8.591&0.91          &12.012a &0.52 \\ \hline
8.726&1.00          &12.012b &0.73 \\ \hline  
9.246&0.28          &12.230 &0.032 \\ \hline  
9.262&0.85          &12.500 &0.27 \\   \hline  
9.754&0.39          &12.654 & 0.88 \\ \hline  
9.904&0.99          &12.795 &0.76 \\ \hline  
9.984&0.20          &12.806 &0.42 \\ \hline  
10.358&0.40         &12.897a &0.87  \\ \hline        
10.686a&0.76        &12.897b &$6.9 \times 10^{-4}$
\end{tabular}\caption{Eigenvalues $k$ and the corresponding 
significance levels $\alpha_D$ for the test of the hypothesis
 $I_N(x) \neq I(x)$ for the Thurston manifold. The injectivity radius 
is maximal at the base point.} 
\label{tab:KScenter}
\end{center} 
\end{table}
The agreement with the RMT prediction is
fairly good for most of eigenmodes which is consistent with the
previous computation in \cite{Inoue1}. However, for five degenerated modes, 
the non-Gaussian signatures are prominent (in \cite{Inoue1}, two modes 
in ($k\!<\!10$) have been missed). Where does this non-Gaussianity come
from? 
\\
\indent
First of all, we must pay attention to the fact that the
expansion coefficients $a_{\nu l m}$ depend on the observing point.
In mathematical literature the point is called the \textit{base point}. 
For a given base point, it is possible to construct a particular
class of fundamental domain called the \textit{Dirichlet}
(\textit{fundamental}) \textit{domain} which is a convex polyhedron.
A Dirichlet domain $\Omega(x)$ centered at a base point $x$ is defined as
\BE
\Omega(x)=\bigcap_{g} H(g,x)~~~,H(g,x)=\{z|d(z,x)<d(g(z),x)\},
\EE
where $g$ is an element of a Kleinian group $\Gamma$(a discrete isometry
 group of $PSL(2,{\mathbb C})$) 
and $d(z,x)$ is the proper distance between $z$ and $x$. 
\\
\indent 
The shape of the Dirichlet domain depends on the base point but the
volume is invariant. Although the base point can be chosen arbitrarily, 
it is a standard to choose a point $Q$ 
where the injective radius \footnote{The injective radius of a point
$p$ is equal to half the length of the shortest periodic geodesic at
$p$.} is locally maximal. More intuitively, $Q$ is a center where one
can put a largest connected ball on the manifold.  
If one chooses other point as the base point, the nearest copy of the base
point can be much nearer. The reason to choose $Q$ as a base point is 
that one can expect the corresponding Dirichlet domain to have 
many symmetries at $Q$ \cite{Weeks2}.
\\
\indent
\begin{figure}[tpb]
\centerline{\psfig{figure=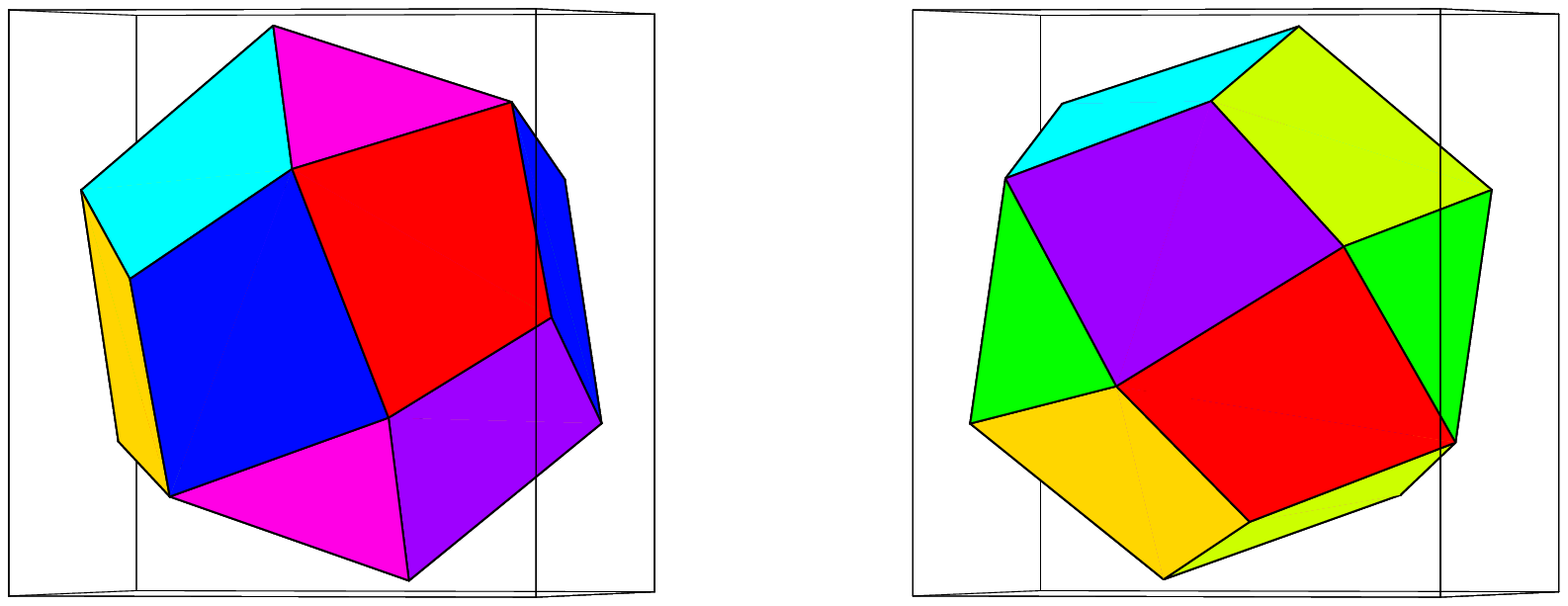,width=17cm}}
\caption{A Dirichlet domain of the Thurston manifold
in the Klein coordinates viewed from opposite directions at $Q$
where the injectivity radius is locally maximal. The Dirichlet domain
has a $Z2$ symmetry(invariant by $\pi$-rotation)at $Q$.}
\label{fig:SYMFDT}
\end{figure}
As shown in Fig.4, the Dirichlet domain at $Q$ has 
a $Z2$ symmetry (invariant by $\pi$-rotation) 
if all the congruent faces are identified. Generally, 
congruent faces 
are distinguished but it is found that these five modes have exactly the same
values of eigenmodes on these congruent faces. Then one can no longer
consider $a_{\nu l m}$'s as ''independent'' random numbers. Choosing    
the invariant axis by the $\pi$-rotation as the $z$-axis,
$a_{\nu l m}$'s are zero for odd $m$'s which leads
to the observed non-Gaussian behavior.
It should be noted that the observed $Z2$ symmetry is not the subgroup
of the isometry group (or \textit{symmetry group} in mathematical literature)
$D2$ (dihedral group with order $2$) of the Thurston manifold 
since the congruent faces must be actually distinguished in the 
manifold\footnote{The observed $Z2$
symmetry is considered to be a ``hidden symmetry'' which is a symmetry
of the finite sheeted cover of the manifold (which tessellates the
manifold as well as the universal covering space). 
For instance, the Dirichlet domain of the Thurston manifold
can be tessellated by four pieces with three neighboring
kite-like quadrilateral faces and one equilateral triangle 
on the boundary and seven faces which contain the center as a vertex. 
By identifying the four pieces (by a tetrahedral symmetry), one obtains an
orbifold which has a $Z2$ symmetry. } 
\\
\indent 
Thus the observed non-Gaussianity is caused by a particular choice of the
base point. However, in general, the chance that we actually observe any
symmetries (elements of the isometry group of the manifold or the finite
sheeted cover of the manifold) is expected to be 
very low. Because a fixed point by an element of the isometric group
is either a part of 1-dimensional line (for instance, an axis of a
rotation) or an isolated point (for instance, a center of an antipodal map). 
\\
\indent
In order to confirm that the chance is actually low, the KS statistics
$\alpha_D$ of $a_{\nu l m}$'s are computed at 300 base points which are 
randomly chosen. \begin{table}
\begin{center}
\begin{tabular}{cccp{22mm}|cccc}   
\multicolumn{4}{c|}{Weeks}&\multicolumn{4}{c}{Thurston} 
\\ \hline
\multicolumn{1}{c}{k} &
\multicolumn{1}{c}{$<\alpha_D>$} &
\multicolumn{1}{c}{k} &
\multicolumn{1}{p{22mm}|}{$<\alpha_D>$} &
\multicolumn{1}{c}{k} &
\multicolumn{1}{c}{$<\alpha_D>$} &
\multicolumn{1}{c}{k} &
\multicolumn{1}{c}{$<\alpha_D>$} 
\\ \hline
5.268&0.58& 10.452b &~~0.62&  5.404&0.63 &10.686b& 0.62 \\ \hline
5.737a&0.61& 10.804&~~0.63&5.783&0.61 &10.737 & 0.62 \\ \hline
5.737b&0.61& 10.857&~~0.62&6.807a &0.62 &10.830 &  0.63\\ \hline
6.563&0.62 & 11.283&~~0.57&6.807b &0.62 &11.103a &0.59 \\ \hline
7.717&0.59&  11.515&~~0.61&6.880&0.63 &11.103b &0.60\\ \hline
8.162&0.61&  11.726a&~~0.63&7.118&0.61 &11.402 & 0.61 \\ \hline
8.207a&0.65& 11.726b&~~0.59&7.686a&0.61 &11.710 & 0.62 \\ \hline
8.207b&0.61& 11.726c&~~0.61&7.686b&0.63 &11.728 &0.64 \\ \hline
8.335a&0.59& 11.726d&~~0.61&8.294&0.60  &11.824 &0.62  \\ \hline
8,335b&0.62& 12.031a&~~0.60&8.591&0.60  &12.012a &0.63 \\ \hline
9.187&0.59&  12.031b&~~0.60&8.726&0.60  &12.012b &0.61 \\ \hline  
9.514&0.56&  12.222a&~~0.61&9.246&0.60  &12.230 &0.60 \\ \hline  
9.687&0.61&  12.222b&~~0.62&9.262&0.63  &12.500 &0.63 \\   \hline  
9.881a&0.61& 12.648&~~0.59&9.754&0.62  &12.654 & 0.62 \\ \hline  
9,881b&0.62& 12.789&~~0.59&9.904&0.60  &12.795 &0.62\\ \hline  
10.335a&0.63&  & &          9.984&0.60  &12.806 &0.62 \\ \hline  
10.335b&0.60&  & &          10.358&0.62 &12.897a &0.62  \\ \hline        
10.452a&0.63&  & &           10.686a&0.60 &12.897b &0.56
\end{tabular}\caption{Eigenvalues $k$ and corresponding 
averaged significance levels $<\alpha_D>$ based on 300 realizations
of the base points for the test of the hypothesis
 $I_N(x) \neq I(x)$ for the Weeks and the Thurston manifolds.} 
\label{tab:KS}
\end{center} 
\end{table}
As shown in table 2, the averaged significance levels $<\!\alpha_D\!>$
are remarkably consistent with the 
Gaussian prediction. $1\sigma$ 
of $\alpha_D$ are found to be 0.26 to 0.30. 
\\
\indent
Next, we apply the run test for testing the randomness of $a_{\nu l
m}$'s where each set of $a_{\nu l m}$ 's are ordered as $l(l+1)+m+1$ 
(see \cite{Hog}). Suppose that we have $n$ observations
of the random variable $U$ which falls above the median and n 
observations of the random variable
$L$ which falls below the median. 
The combination of those variables into $2 n$ observations
placed in ascending order of magnitude yields
\begin{center}
\large\textit{
\U{UUU} \U{LL} \U{UU} \U{LLL} \U{U} \U{L} \U{UU} \U{LL}},
\end{center}
Each underlined group which consists of successive 
values of $U$ or $L$ is called \textit{run}. The total number of 
run is called the \textit{run number}.
The run test is useful because the run number 
always obeys the Gaussian statistics in the limit
$n\!\rightarrow\!\infty$ regardless of the type of the
distribution function of the random variables.
\begin{table}
\begin{center}
\begin{tabular}{cccp{22mm}|cccc}   
\multicolumn{4}{c|}{Weeks}&\multicolumn{4}{c}{Thurston} 
\\ \hline
\multicolumn{1}{c}{k} &
\multicolumn{1}{c}{$<\alpha_r>$} &
\multicolumn{1}{c}{k} &
\multicolumn{1}{p{22mm}|}{$<\alpha_r>$} &
\multicolumn{1}{c}{k} &
\multicolumn{1}{c}{$<\alpha_r>$} &
\multicolumn{1}{c}{k} &
\multicolumn{1}{c}{$<\alpha_r>$} 
\\ \hline
5.268&0.51& 10.452b &~~0.52&  5.404&0.48 &10.686b& 0.51 \\ \hline
5.737a&0.48& 10.804&~~0.52&5.783&0.45 &10.737 & 0.49 \\ \hline
5.737b&0.45& 10.857&~~0.53&6.807a &0.53 &10.830 &0.53\\ \hline
6.563&0.54 & 11.283&~~0.49&6.807b &0.50 &11.103a &0.52 \\ \hline
7.717&0.50&  11.515&~~0.51&6.880&0.47 &11.103b &0.53\\ \hline
8.162&0.54&  11.726a&~~0.51&7.118&0.50 &11.402 & 0.51 \\ \hline
8.207a&0.52& 11.726b&~~0.48&7.686a&0.49 &11.710 & 0.51 \\ \hline
8.207b&0.49& 11.726c&~~0.49&7.686b&0.52 &11.728 &0.49 \\ \hline
8.335a&0.53& 11.726d&~~0.48&8.294&0.50 &11.824 &0.54  \\ \hline
8,335b&0.50& 12.031a&~~0.54&8.591&0.50  &12.012a &0.51 \\ \hline
9.187&0.53&  12.031b&~~0.51&8.726&0.51  &12.012b &0.49 \\ \hline  
9.514&0.55&  12.222a&~~0.54&9.246&0.43  &12.230 &0.51 \\ \hline  
9.687&0.53&  12.222b&~~0.50&9.262&0.50  &12.500 &0.48 \\   \hline  
9.881a&0.51& 12.648&~~0.54&9.754&0.54 &12.654 & 0.48 \\ \hline  
9,881b&0.51& 12.789&~~0.48&9.904&0.52  &12.795 &0.50\\ \hline  
10.335a&0.54&  & &          9.984&0.49  &12.806 &0.51 \\ \hline  
10.335b&0.51&  & &          10.358&0.53 &12.897a &0.57  \\ \hline        
10.452a&0.53&  & &           10.686a&0.51 &12.897b &0.55
\end{tabular}\caption{Eigenvalues $k$ and corresponding 
averaged significance levels $<\!\alpha_r\!>$ for the test of the hypothesis
that the $a_{\nu l m}$'s are not random numbers for the Weeks and
Thurston manifolds. $\alpha_r$'s at 300 points which are randomly
chosen are used for the computation.} 
\label{tab:KS}
\end{center} 
\end{table}
As shown in table 3, averaged significance levels  $<\!\alpha_r\!>$ are
very high (1$\sigma$ is 0.25 to 0.31). 
Thus each set of $a_{\nu l m}$'s ordered as $l(l+1)+m+1$ can be
interpreted as a set of Gaussian pseudo-random numbers except for limited
choices of the base point where one can observe symmetries of 
eigenmodes. 
\\
\indent
Up to now, we have considered $l$ and $m$ as the index numbers of $a_{\nu
l m}$ at a fixed base point. However, for a fixed $(l,m)$, the
statistical property of a set of $a_{\nu l m}$'s at a number of 
different base points is also important since the temperature 
fluctuations must be averaged all over the places 
for spatially inhomogeneous models. 
From Fig.5, one can see the behavior of m-averaged significance 
levels  
\BE
\alpha_D(\nu, l)\equiv \sum_{m=-l}^{l} \f{\alpha_D(a_{\nu l m})}{2l+1}
\EE
which are calculated based on 300 realizations of the base points.  
It should be noted that each $a_{\nu l m}$ at a particular
base point is now considered to be ''one realization'' whereas
a choice of $l$ and $m$ is considered to be ''one realization'' in
the previous analysis (table 1).
The agreement with the RMT prediction is considerably good for 
components $l\!>\!1$. For components $l\!=\!1$, the disagreement
occurs for only several modes. However, the non-Gaussian 
behavior is distinct in $l\!=\!0$ components. 
What is the reason of the non-Gaussian 
behavior for $l\!=\!0$? 
\begin{figure}[tpb]
\centerline{\psfig{figure=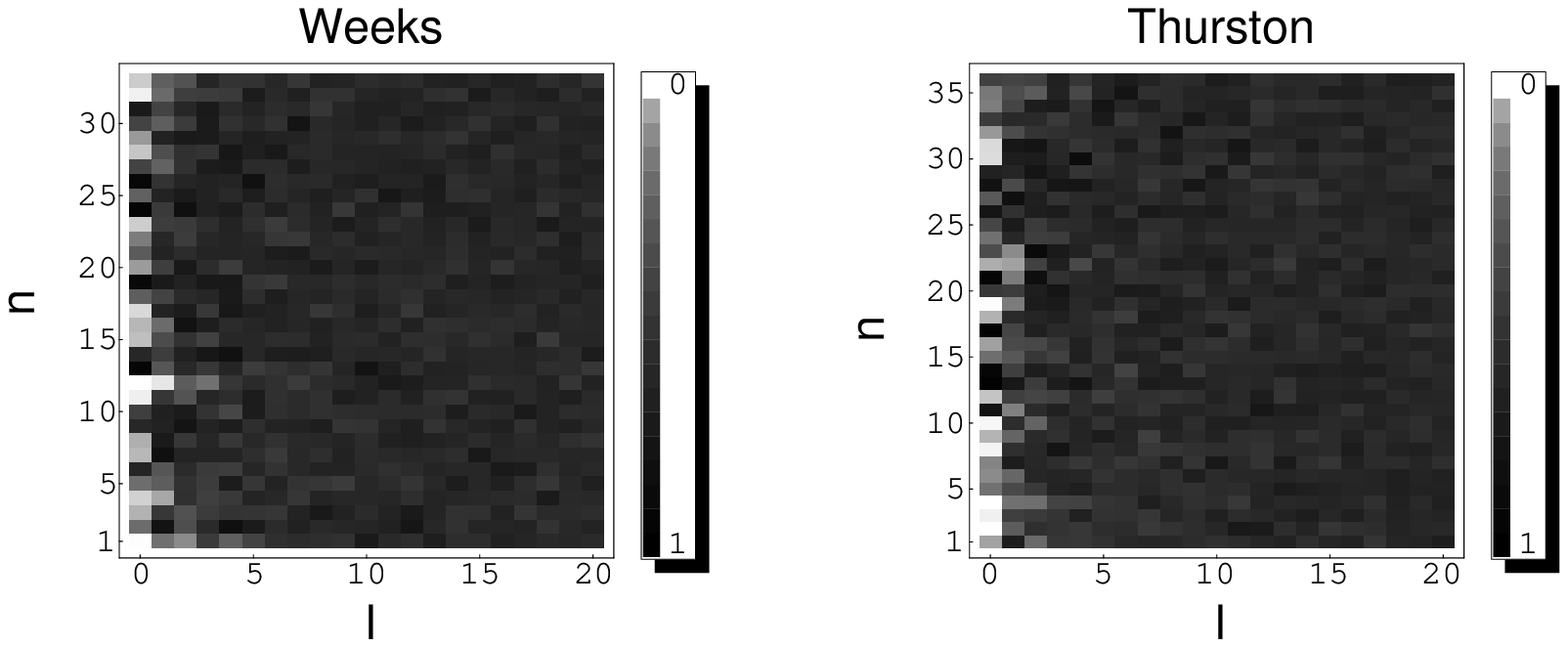,width=17cm}}
\caption{Plots of m-averaged significance 
levels $\alpha_D(\nu, l)$ 
based on 300 realizations for the Weeks and the Thurston manifolds
($0 \! \le\! l \!\le\! 20$ and $k<13$). $n$ denotes the index number
which corresponds to an eigenmode $u_k$ where the number of eigenmodes 
less than $k$ is equal to $n$ ($k(n\!=\!1)$ is the lowest non-zero
eigenvalue). The accompanying palettes show the 
correspondence between the level of the grey and the value.}
\label{fig:SYMFDT}
\end{figure}
Let us estimate the values of the expansion coefficients for
$l\!=\!0$. 
In general, the complex expansion coefficients $\xi_{\nu l m}$ 
can be written as,
\BE
\xi_{\nu l m}(\chi_0)=\f{1}{X_{\nu l}(\chi_0)} 
\int u_\nu(\chi_0,\theta,\phi)\, 
Y^*_{l m}(\theta,\phi) d \Omega. \label{eq:xi}
\EE 
For $l\!=\!0$, the equation becomes
\BE
\xi_{\nu 0 0}(\chi_0)=-\f{i}{2\sqrt{2}} \f{\sinh{\chi_0}}{\sin{\nu
 \chi_0}} \int u_\nu(\chi_0,\theta,\phi)\, d \Omega. \label{eq:xi0}
\EE 
Taking the limit $\chi_0 \rightarrow 0$, one obtains,
\BE
\xi_{\nu 0 0}=-\f{2 \pi u_\nu(0) i}{\nu}.
\EE
Thus $a_{\nu 0 0}$ can be written in terms of the value of the
eigenmode at the base point. 
As shown in Fig.1, the lowest eigenmodes have only 
one ''wave'' on scale of the topological identification scale $L$ 
(which will be defined later on) 
inside a single Dirichlet domain which implies that the random
behavior within the domain may be not present. 
Therefore, for low-lying eigenmodes,
one would generally expect non-Gaussianity in a set of 
$a_{\nu 0 0}$ 's. 
However, for high-lying eigenmodes, this may not be the case 
since these modes have a number of ''waves'' on scale of $L$ 
and they may change their values locally in a almost random fashion.
\\
\indent
The above argument cannot be applicable to $a_{\nu l m}$ 's for 
$l \!\neq\! 0$ where ${X_{\nu l}}$ approaches zero in the limit 
$\chi_0\!\rightarrow\! 0$ while the integral term
\BE
 \int u_\nu(\chi_0,\theta,\phi)\, 
Y^*_{l m}(\theta,\phi) d \Omega 
\EE 
also goes to zero because of the symmetric property of the spherical
harmonics. Therefore $a_{\nu l m}$ 's  cannot be written in terms of
the local value of the eigenmode for $l \!\neq\! 0$. 
For these modes, it is better to consider the opposite
limit $\chi_0 \rightarrow \infty$. It is numerically found that 
the sphere with very large radius $\chi_0$ intersects each copy of 
the Dirichlet domain almost randomly (the 
pulled back surface into a single Dirichlet domain chaotically
fills up the domain). Then the values of the eigenmodes
on the sphere with very large radius vary in an almost random
fashion. For large $\chi_0$, we have 
\BE
X_{\nu l}(\chi_0)\!\propto\!e^{-2
\chi_0+\phi(\nu, l)i},
\EE
where $\phi(\nu, l)$ describes the phase factor. Therefore, the order
of the integrand in Eq. (\ref{eq:xi}) is approximately $e^{-2
\chi_0}$ since Eq. (\ref{eq:xi}) does not depend on the choice of
$\chi_0$. As the spherical harmonics do not have correlation with 
the eigenmode $u_{\nu}(\chi_0,\theta,\phi)$, the integrand varies
almost randomly for different choices of $(l,m)$ or base points. 
Thus we conjecture that Gaussianity of $a_{\nu l m}$'s have their origins 
in the chaotic property of the sphere with large radius in CH spaces. 
The property may be related to the 
classical chaos in geodesic flows\footnote{If one considers a great
circle on a sphere with large radius, the length of the circle is
very long except for rare cases in which the circle ``comes back'' 
before it wraps around in the universal covering space. Because the 
long geodesics in CH spaces chaotically (with no particular direction 
and position) wrap through the manifold, it is natural to assume that 
the great circles also have this chaotic property.}  
. 
\\
\indent
So far we have seen the Gaussian pseudo-randomness 
of the $a_{\nu l m}$'s. Let us now consider the statistical properties
of the expansion coefficients.
As the eigenmodes have oscillatory features, it is natural to expect
that the averages are equal to zero. In fact, the averages of
$<\!a_{\nu l m}\!>$ 's over $0\!\le\! l\!\le\! 20$ and $-l\!\le\!
m\!\le\! -l$ and 300 realizations of base points for each $\nu$-mode
are numerically found to be $0.006\pm 0.04-0.02$(1$\sigma$) 
for the Weeks manifold, 
and $0.003\pm 0.04-0.02$(1$\sigma$) for the Thurston
manifold. Let us next consider the $\nu$-dependence ($k$-dependence) 
of the variances $Var(a_{\nu l m})$. In order to crudely 
estimate the $\nu$-dependence, 
we need the angular size $\delta \theta $
of the characteristic length of the eigenmode $u_{\nu}$
at $\chi_0$\cite{Inoue1}
\BE
\delta \theta^2\approx
\f{16 \pi^2~V\!o\,l(M)}  
{k^2 (\sinh(2(\chi_o+r_{ave}))-\sinh(2(\chi_o-r_{ave}))-4 r_{ave})},
\label{eq:deltheta}
\EE
where $V\!o\,l(M)$ denotes the volume of a manifold $M$ and $r_{ave}$
is the averaged radius of the Dirichlet domain. There is an arbitrariness
in the definition of $r_{ave}$. Here we define $r_{ave}$ as the radius 
of a sphere with volume equivalent to
the volume of the manifold, 
\BE
V\!o\,l(M)=\pi(\sinh(2 r_{ave})-2 r_{ave}),
\EE
which does not depend on the choice of the base point.
The topological identification length $L$ is defined as $L\!=\!2r_{ave}$.
For the Weeks and the Thurston manifold, $L\!=\!1.19$ and 
$L\!=\!1.20$ respectively. From
Eq. (\ref{eq:deltheta}), for large $\chi_0$, one can approximate
$u_{\nu}(\chi_o)\sim u_\nu'(\chi'_o)$ by choosing an appropriate
radius $\chi'_o$ which satisfies 
$\nu^{-2}\exp(-2\chi_o)=\nu'^{-2}\exp(-2\chi'_o)$. 
Averaging Eq. (\ref{eq:xi}) over $l$'s and $m$'s or the base
points, for large $\chi_0$, one obtains,
\BE
\bigl <|\xi_{\nu' l m} |^2 \bigr >
\sim 
\f{\exp(-2\chi_o)}{\exp(-2\chi'_o)}
\bigl<|\xi_{\nu l m} |^2 \bigr >,
\EE
which gives $\bigl<|\xi_{\nu l m}|^2 \bigr> \sim \nu^{-2}$.
Thus the variance of $a_{\nu l m}$'s is proportional to $\nu^{-2}$.
The numerical results for the two CH manifolds shown in Fig.6
clearly support the $\nu^{-2}$ dependence of the variance. 
\BF[tpb]
\centerline{\psfig{figure=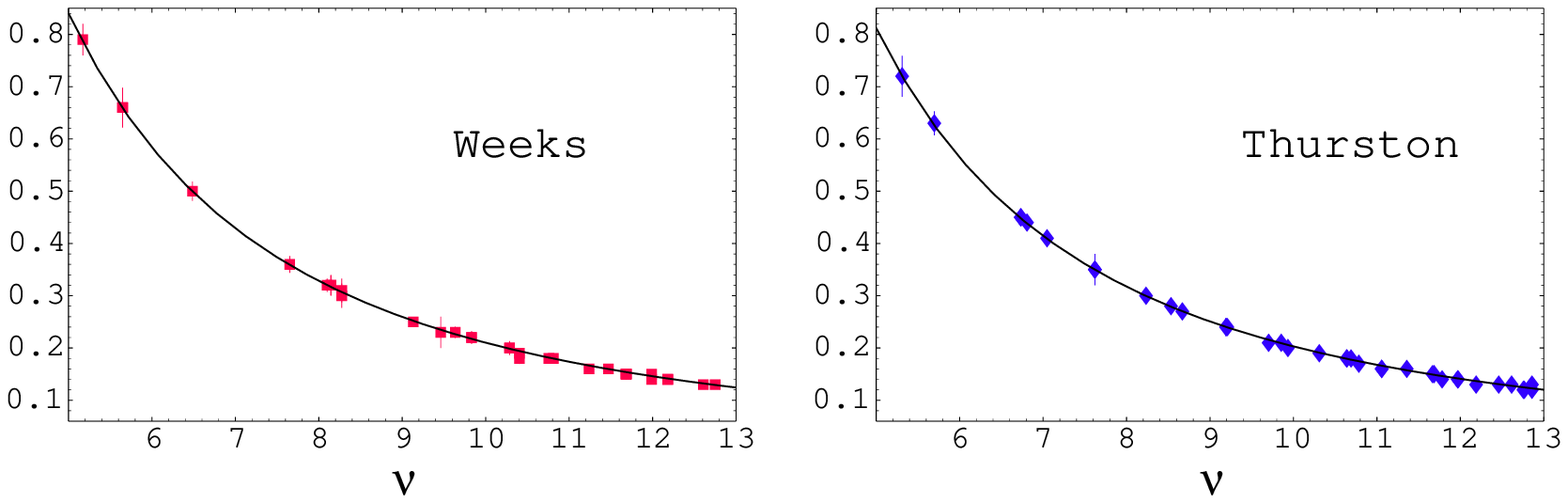,width=17cm}}
\caption{Averaged squared $a_{\nu}$'s ($k\!<\!13$)
based on 300 realizations of the base points for the Weeks and the Thurston
manifold with $\pm 1\sigma$ run-to-run variations.  
 $a_{\nu}$ is defined to be $Var(a_{\nu l m})$ averaged over $0\! \le\! l\!
\le\!20$ and $-l\! \le\! m\!\le\!l$. 
The best-fit curves for the Weeks and the Thurston manifolds are 
$21.0\nu^{-2}$ and $20.3\nu^{-2}$, respectively.} 
\label{fig:SEC}
\EF
\\
\indent
As we have seen, the property of eigenmodes on general CH manifolds is
summarized in the following conjecture:
\\
\\
\textit{Conjecture: Except for the base points which are too close to any
fixed points by symmetries, for a fixed $\nu$, a set of the expansion 
coefficients $a_{\nu l m}$ over $(l,m)$'s can be considered as 
Gaussian pseudo-random numbers. For a fixed $(\nu l m)~(l>0)$, 
the expansion coefficients at different base points that are randomly chosen 
can also be considered as Gaussian pseudo-random numbers. In either
case, the variance is proportional to $\nu^{-2}$ and the average is zero.}    
\section{NON-GAUSSIANITY IN OBSERVABLE ANGULAR POWER SPECTRA}
As mentioned in the last section, perturbations in CH models
are written in terms of linear combinations of eigenmodes 
and the time evolution of the perturbations.
Because the time evolution of the perturbations coincides with that
in open models, once the expansion coefficients $\xi_{\nu l m}$ (or 
$a_{\nu l m}$) are given, the evolution of perturbations in 
CH models can be readily obtained.
\\
\indent
If one assumes that the perturbation is  
a adiabatic scalar type without anisotropic pressure, 
and the subhorizon effects
such as acoustic oscillations of the temperature and the velocity
of the bulk fluid, and the effect of the radiation contribution
at high z are negligible, the time evolution of the 
growing mode of the Newtonian curvature $\Phi$ is analytically given
as (see e.g. \cite{Kodama,Mukhanov})
\begin{equation}
\Phi(\eta)=
\f{5(\sinh^2 \eta-3 \eta\sinh\eta+4 \cosh\eta-4)}
{(\cosh\eta-1)^3},
\end {equation}  
where $\eta$ denotes the conformal time. In terms of $\Phi$, 
the temperature fluctuation in the sky are written as
\BEA
\f{\Delta T(\n)}{T}&=&\sum_{lm}a_{lm}Y_{lm}(\n)
\nonumber
\\
&=&\sum_{\nu l m} \Phi_{\nu}(0)\xi_{\nu l m}F_{\nu l
}(\eta_0)Y_{lm}(\n), \label{eq:deltaT}
\EEA
where
\BE
F_{\nu l}(\eta_0)
\!\!\equiv -\f{1}{3}
\Phi(\eta_\ast) X_{\nu l}(\eta_0\!-\!\eta_\ast)
\!-\!\! 2 \!\!\int_{\eta_\ast}^
{\eta_0}\!\!\!\!\!\!d \eta\, 
\f{d\Phi}{d \eta}X_{\nu l}(\eta_0\!-\!\eta).\label{eq:cor}
\EE 
Here $\Phi_{\nu}(0)$ is the initial value of the curvature
perturbation and $\eta_\ast$ and $\eta_0$ 
are the conformal time of the last scattering and the present 
conformal time, respectively. The angular power spectrum $ 
C_l $ is defined as
\BEA
(2\,l+1)\, C_l
&=&\sum_{m=-l}^{l} \langle |a_{lm}|^2 \rangle
\nonumber
\\
&=&\sum_{\nu,m}\f{4 \pi^4~{\cal P}_\Phi(\nu) }
{\nu(\nu^2+1)\textrm{Vol}(M)}~\langle |\xi_{\nu l m}|^2 \rangle 
|F_{\nu l}(\eta_0)|^2, 
\EEA
where ${\cal P}_\Phi(\nu) $ is the initial power spectrum. It should
be noted that the above formula converges to that of open models
in the short-wavelength limit (summation to integration) 
provided that $<\!|\xi_{\nu l m}|^2\!>$ 
is proportional to $\nu^{-2}$. The reason is as follows: Let us denote 
the number of eigenmodes with eigenvalues equal to or less than
$\nu$ by $N(\nu)$. In the short-wavelength limit $\nu\!>>\!1$ 
one can use Weyl's 
asymptotic formula which leads to
\BE
\f{d N(\nu)}{d\nu}=\f{\textrm{Vol}(M)}{2 \pi^2}\nu^2. \label{eq:Weyl}
\EE 
Thus the $\nu^2$ dependence in Eq.(\ref{eq:Weyl}) is exactly cancelled out 
by the $\nu^{-2}$ dependence of eigenmodes. In what follows we assume 
the extended Harrison-Zel'dovich 
spectrum, \textit{i.e.} ${\cal P}_\Phi(\nu)\!=\!Const.$
(in the flat limit, it converges to the
scale invariant Harrison-Zel'dovich spectrum) as the
initial power spectrum.
\\
\indent
In estimating the temperature correlations, the non-diagonal
terms ($l\!\neq\!l'$ or $m\!\neq\! m'$) may
not be negligible if the background spatial hypersurface is not
isotropic, in other words, the angular power spectrum $C_l$
may not be sufficient in describing the temperature correlations
since $C_l$ provides us with only an isotropic information of 
statistics of the correlations.
However, this is not the case for CH models to which the conjecture
proposed in Sec.~II is applicable.
Based on the Copernican principle,
it is not likely that we are at the center of any symmetries. 
Therefore, in order to statistically estimate the temperature 
correlations in the globally inhomogeneous background space, 
one has to consider an ensemble of fluctuations with
different initial conditions at different places (or base points) 
with different orientations.  
Almost all the anisotropic
information is lost in the spatial averaging process since
the eigenmodes are Gaussian.
\\
\begin{figure}[tpb]
\centerline{\psfig{figure=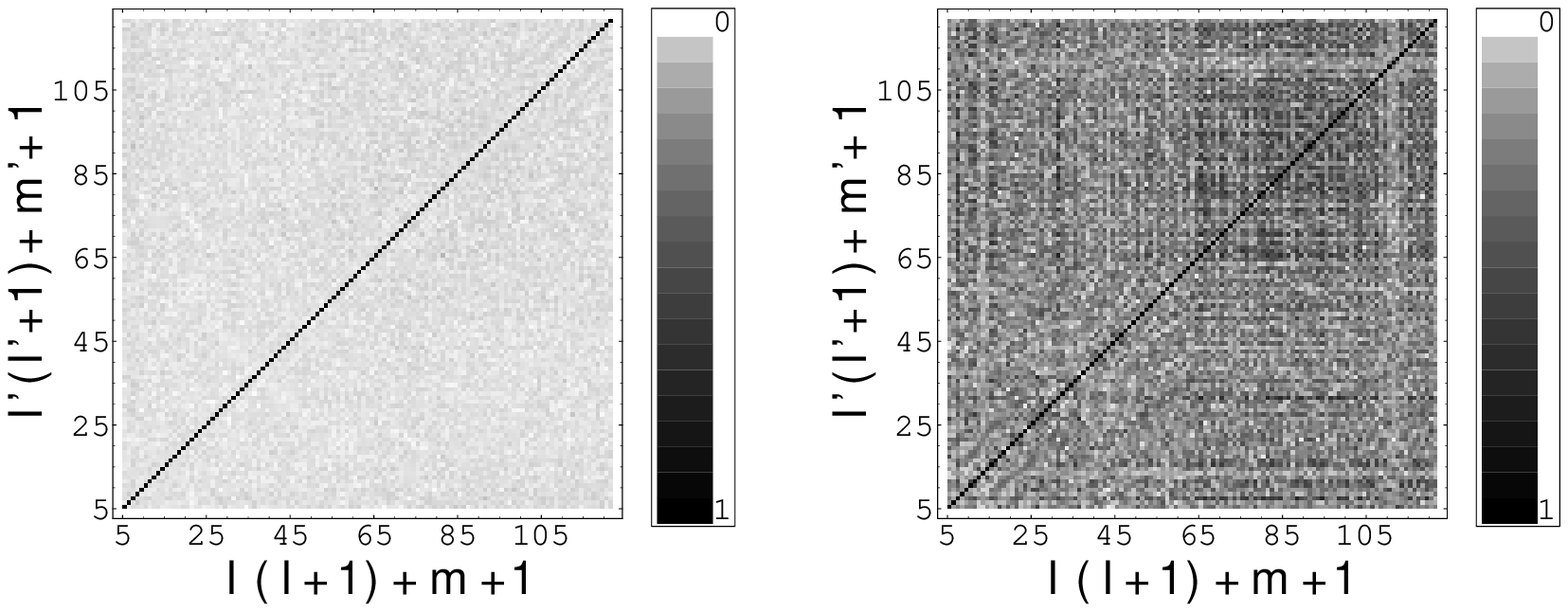,width=17cm}}
\caption{Contributions of non-diagonal terms in the temperature correlations 
in unit of diagonal terms which are defined as 
$f^{lm}_{l'm'}\!=\!|\!<\!a_{lm}a\ast_{l'm'}\!>\!|
~/\sqrt{<|a_{lm}|^2><|a_{l'm'}|^2>}$ for the Thurston model with
$\Omega_0$=0.3. The four-dimensional space ($l,m,l',m'$) is
represented in the two-dimensional space as 
$(n,n')\!=\!(l(l+1)+m+1,l'(l'+1)+m'+1)$ for $2\!\le\!l\!\le\!10,
-l\!\le\!m\!\le\!l$ and  $2\!\le\!l'\!\le\!10,
-l'\!\le\!m'\!\le\!l'$.  $f^{lm}_{l'm'}$'s
are represented by the level of grey shown in the accompanying
palettes. The left figure represents $f^{lm}_{l'm'}$'s averaged over
300 realizations of the base points with infinite number of 
initial conditions for the Newtonian curvature. The right figure 
represents $f^{lm}_{l'm'}$'s at a base point where the injective
radius is maximal with infinite number of 
initial conditions. The computation is based on 36
eigenmodes($k\!<\!13$) that are numerically obtained by using the
direct boundary element method. The averaged values of the  
non-diagonal $f^{lm}_{l'm'}$'s ($l\! \neq\! l'$ or $m\! \neq\! m'$)are
0.016(left) and 0.25(right).} 
\label{eq:NonDiaT}
\end{figure}
\indent
As shown in Fig.7, for 300 realizations of observing points(left), 
the averaged absolute values of the off-diagonal elements 
in unit of diagonal elements are very small ($\sim0.016$)
whereas their contributions seem to be not negligible ($\sim0.25$) at one
particular observing point(right) where one can observe a symmetry of
the Dirichlet domain. Thus the statistical property of
the temperature correlation can be estimated by using $C_l$'s provided 
that the eigenmodes are Gaussian which validates the previous analyses
using $C_l$'s for constraining the CH models 
\cite{Aurich3,Inoue2,Inoue3,Cornish1}. The
spatial averaging process\footnote{In general, one should include 
an averaging process over different choices of orientation of
coordinates as well as  
an averaging process over different choices of the observing point. 
Nevertheless, the Gaussian conjecture in Sec.~II implies that the 
eigenmodes on CH
spaces are ``$SO(3)$ invariant'' \cite{Mag1}
if averaged all over the space. Therefore, omission of the 
averaging procedure for different orientations of coordinates 
make no difference.}
must be taken into account since
there is no reason to believe that we are in the center of any symmetries.
\\
\indent
If the initial conditions satisfy 
$(\Phi_{\nu}(0))^{-2}\!\propto\!\nu(\nu^2+1)$ 
that corresponds to the extended 
Harrison-Zel'dovich spectrum
, then Eq.(\ref{eq:deltaT}) tells us that the temperature fluctuation
is Gaussian since it is equal to a sum
of Gaussian (pseudo-)random numbers at almost all the observing points. 
In this case, the Gaussian randomness of the temperature fluctuations 
in CH models can be solely attributed to the geometrical property of 
the space (geometric Gaussianity) which may be related to the
deterministic chaos of the corresponding classical system. In other
words, the Gaussian randomness can be
explained in terms of the classical physical quantities 
without considering the initial quantum fluctuations provided that the above
conditions are initially (deterministically) satisfied.
\\
\indent
However, it is much natural to assume that $\Phi_{\nu}(0)$'s are
also random Gaussian as
in the inflationary scenarios in which Gaussianity 
(on large scales) of 
the temperature fluctuations has its origin in Gaussianity of 
the initial quantum fluctuations because the angular powers are
generally similar
to the extended Harisson-Zel'dovich spectrum.  Then the statistical
properties of the temperature fluctuations are determined by the sum of the 
products of the two independent Gaussian random 
numbers (the initial fluctuations and
the expansion coefficients of the eigenmodes).
\\
\indent
Let us calculate the distribution function $F(Z,\sigma_Z)$ 
of a product of two independent random numbers $X$ and $Y$ 
that obey the Gaussian (normal) distributions 
$N(X;0,\sigma_X)$ and $N(Y;0,\sigma_Y)$, respectively.
\BE
N(X;\mu,\sigma)\equiv \f{1}{\sqrt{2 \pi} \sigma}
\textrm{e}^{-(X-\mu)^2/2\sigma^2}.
\EE
Then $F(Z\!=\!XY,\sigma_Z)$ is readily given by
\BEA
F(Z,\sigma_Z)&=&2\int_0^\infty N(Z/Y,0,\sigma_X)N(Y,0,\sigma_Y)
\f{dY}{Y}
\nonumber
\\
&=& \f{1}{\pi \sigma_X \sigma_Y}K_{0}\Bigl ( \f{|Z|}{\sigma_X \sigma_Y}
\Bigr ),
\EEA
where $K_{0}(z)$ is the modified Bessel function. The average of $Z$
is zero and the standard deviation satisfies 
$\sigma_Z=\sigma_X \sigma_Y$. As is well known,
$K_{0}(z)$ is the Green function of the diffusion equation with 
sources distributed along an infinite line.  Although $K_{0}(z)$ is 
diverged at $z\!=\!0$ its integration over $(-\infty,\infty)$ is convergent. 
\begin{figure}[tpb]
\centerline{\psfig{figure=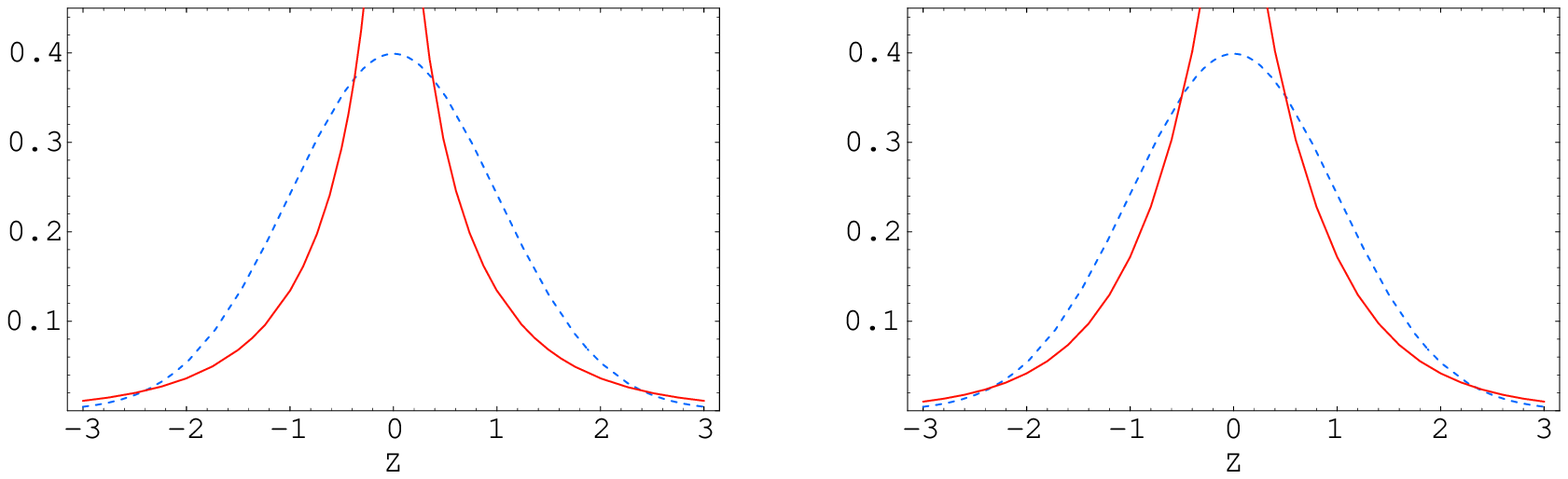,width=17cm}}
\caption{On the left, the distribution function 
$F(Z,1)$ 
for a product of two random Gaussian numbers is plotted in solid
curves. On the right, the distribution function 
$G(Z,1)$ (1$\sigma\!=\!1$) 
of a sum of two random variables that obey  
$F(Z,1/\sqrt{2})$. The dashed curves represent the
Gaussian distribution $N(Z;0,1)$.
}
\label{eq:GtimesG}
\end{figure}
From the asymptotic expansion of the modified
Bessel function
\BE
K_0(z)\sim \sqrt{\f{\pi}{2 z}}\textrm{e}^{-z}\Biggl[1-\f{1^2}{1!8z}
+\f{1^2\cdot 3^2}{2!(8z)^2}-\f{1^2\cdot 3^2\cdot 5^2}{3!(8z)^3}+\ldots
\Biggr],~~~ z>>1,
\EE
one obtains in the lowest order approximation, 
\BE
F(Z,\sigma)\sim \f{1}{\sqrt{2 \pi \sigma
|Z|}}\textrm{e}^{-|Z|/\sigma},~~~Z>>1.
\EE 
Thus $F(Z,\sigma)$ is slowly decreased than the Gaussian distribution 
function with the same variance in the large limit. 
One can see the two non-Gaussian features in
Fig.8(left):the divergence at  $Z\rightarrow 0$ and the 
slow convergence to zero at $Z \rightarrow \infty$.
The slow convergence is an important feature, as we shall see, 
in distinguishing the non-Gaussian models with the Gaussian ones.
In the modest region $0.4\!<\!|Z|\!<\!2.4$,
$F(Z,1)$ is much less than $N(Z,0,1)$. Generally, the temperature fluctuation
is written as a sum of the random variables $Z_i$ which obeys the
distribution function $F(Z_i,\sigma_{Z_{i}})$ for a fixed set of 
cosmological parameters. 
For large-angle fluctuations, only the eigenmodes with large wavelength 
($\!\equiv\!2 \pi/k$)can contribute to the sum. Due to the finiteness of 
the space, the number of eigenmodes which dominantly contribute to 
the sum is finite. Therefore, the fluctuations are
distinctively non-Gaussian. For small-angle fluctuations, 
the number of eigenmodes that contribute to the
sum becomes so large that the distribution function converges to
the Gaussian distribution as the central limit theorem implies.
One can see from Fig.8 (right) that the distribution 
function $G(W,1)$ of $W\!=\!Z1+Z2$ where both $Z1$ and $Z2$ 
obey $F(Z,\sqrt{2})$ is much similar to the Gaussian distribution $N(Z,0,1)$ 
than $F(W,1)$ in the modest region. 
\\
\indent
Now let us see the non-Gaussian features of the observable
angular power spectrum 
$\hat{C_l}$ assuming that the initial fluctuations are Gaussian. 
First of all, we define a statistic 
$\tilde{\chi}^2\! \equiv\! (2l+1)\hat {C_l}/C_l$ where 
\BE
(2l+1)\hat{C_l}\!=\!\sum_{m=-l}^{l}b_{lm}^2.
\EE 
\begin{figure}[tpb]
\centerline{\psfig{figure=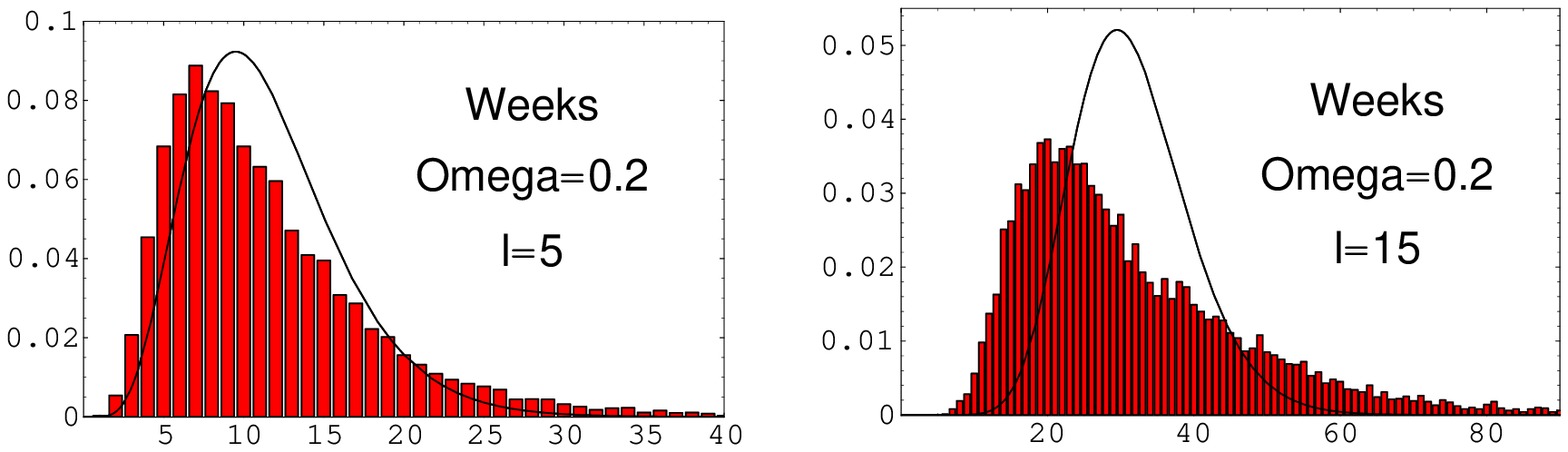,width=17cm}}
\caption{The distributions of $\tilde{\chi}^2\! \equiv\!
(2l+1)\hat{C_l}/C_l$ 
for the Weeks model with $\Omega_0\!=\!0.2$, $l\!=\!5$(left) and $15$(right).  
The horizontal axes represent the values of  $\tilde{\chi}^2$.
The distributions are calculated using 33 eigenmodes ($k<13$)
based on 200 realizations of the 
initial Gaussian fluctuations $\Phi_{\nu}(0)$,
and 200 realizations of the base points. 
The contribution of modes $k\!>\!13$ is approximately
less than 8 percent for $l\!\le\!15$. The solid curves
represent the $\chi^2$ distributions with $11$(left) and $31$(right)  
degrees of freedom.}
\label{eq:GtimesG}
\end{figure}
If the expansion coefficients $b_{lm}$ of the temperature fluctuation in the
sky are Gaussian, $\tilde{\chi}^2$ must obey 
the $\chi^2$ distribution with $2m\!+\!1$ degrees of freedom. 
\begin{figure}[tpb]
\centerline{\psfig{figure=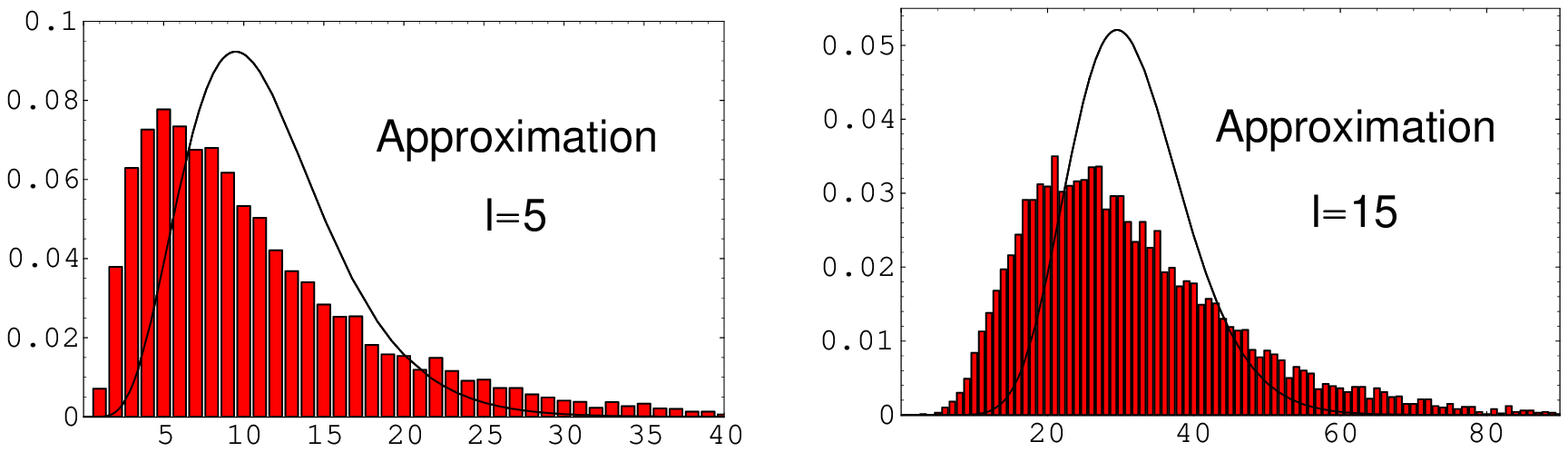,width=17cm}}
\caption{The distributions of $\tilde{\chi}^2\! \equiv\!
(2l+1)\hat{C_l}/C_l$ in an
approximated model in which $b_{lm}$'s obey $G(Z,1)$ for $l\!=\!5$ and 
$l\!=\!15$
based on 40000 realizations for each $b_{lm}$. The horizontal axes
represent the values of $\tilde{\chi}^2$. The solid curves
correspond to the $\chi^2$ distributions with $11$(left) and $31$(right)  
degrees of freedom.}
\label{eq:DISPSEUDO}
\end{figure}
Fig.9 shows the two non-Gaussian features in the distribution of 
$b_{lm}$'s:a slight shift of the peak to the center(zero);
slow convergence to zero for large $\tilde{\chi}^2$.
As shown in Fig.10, the distribution of $\tilde{\chi}^2$ is
approximately obtained by assuming that $b_{lm}$'s obey 
$G(Z,1)$ (actually, the distribution functions of $b_{lm}$'s
are slightly much similar to the Gaussian 
distributions on large angular scales). 
The two non-Gaussian features are attributed to
the nature of the distribution functions of each $b_{lm}$ 
which give large values at $b_{lm}\!\sim\!0$ and decrease 
slowly at $b_{lm}>>1$ compared with the Gaussian distributions. 
\begin{figure}[tpb]
\centerline{\psfig{figure=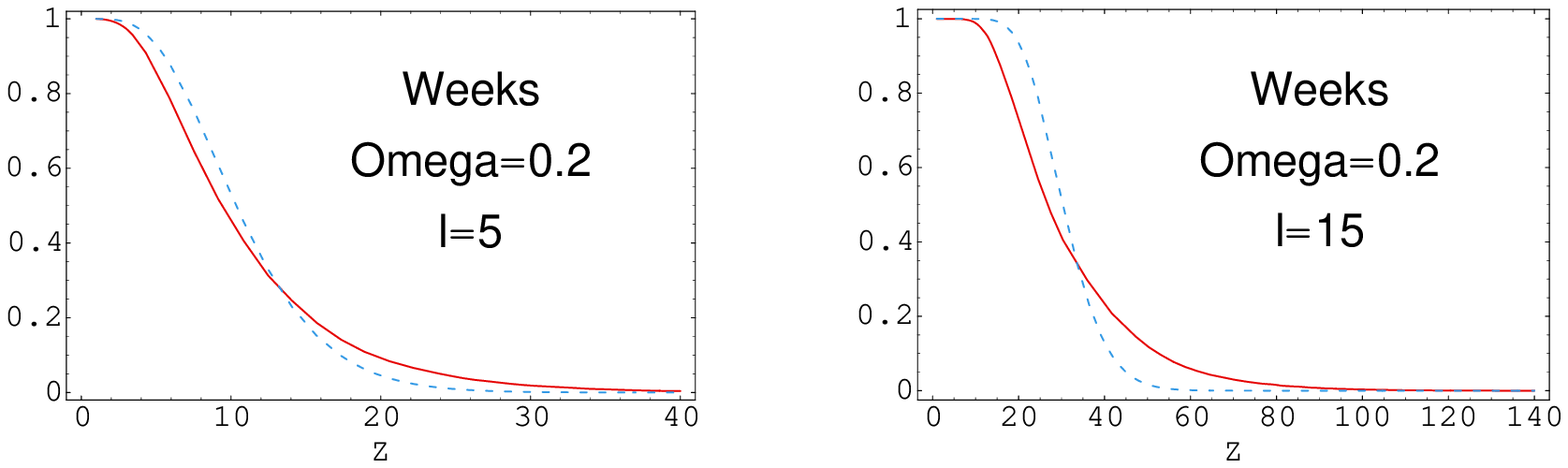,width=17cm}}
\caption{Plots of $1-P(Z)$ ($P(Z)$ is the cumulative 
distribution function) which gives the probability of observing
$X\ge Z$.  The solid curves correspond to  
$1-P(\tilde{\chi}^2)$ for the
Weeks model $\Omega_0\!=\!0.2$, $l\!=\!5$ (left) and $l\!=\!15$ (right). 
The dashed curves correspond to $1-P(\chi^2)$ of the Gaussian model.} 
\label{eq:Prob0.2W}
\end{figure}
\\
\indent
The slow decrease of the distribution of $\tilde{\chi}^2$ 
is important in discriminating the non-Gaussian models with the
Gaussian models. As shown in Fig.11, observing 
$\tilde{\chi}^2\!\sim\!50$ are not improbable for the Weeks $\Omega_0$
model ($l\!=\!15$) whereas it is almost unlikely for the Gaussian model.
Because the distribution is slowly decreased for large
$\tilde{\chi}$, the cosmic variances 
$(\Delta C_l)^2$ 
are expected to be larger than that of the Gaussian models. 
\begin{figure}[tpb]
\centerline{\psfig{figure=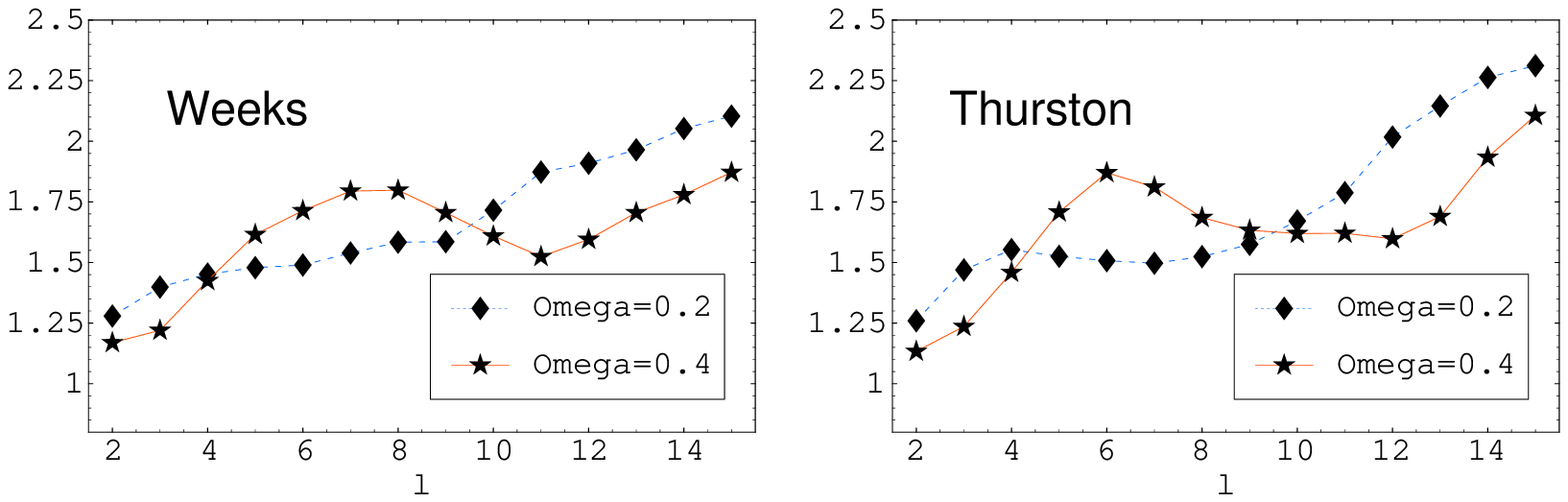,width=18.5cm}}
\caption{Plots of $\Delta C_l(\textrm{CH})/\Delta C_l
(\textrm{Gauss})$ for the two CH models based on 200 realizations
of the initial perturbation $\Phi_\nu(0)$ and 200 realizations of 
the base point.   
$\Delta C_l$ denotes the standard deviation (1$\sigma$) of $\hat{C_l}$.}  
\label{eq:Chioverchi}
\end{figure} 
From Fig.12, on large angular scales($2\!\le\! l \!\le\! 15$), 
one can see that the standard deviations $\Delta C_l$ of $\hat{C_l}$  in
the two CH models are approximately 1-2 times of that for the
Gaussian models. 
\\
\indent
\section{TOPOLOGICAL QUANTITIES}
Topological measures:total area of the excursion regions, total length 
and the genus of the isotemperature contours 
have been used for testing Gaussianity 
of the temperature fluctuations in the COBE DMR
data\cite{Colley,Kogut}. Let us first
summarize the known results for Gaussian fields (see \cite{Gott,Adler}).
\\
\indent
The genus $G$ of the excursion set for
a random temperature field on a connected and simply-connected 2-surface 
can be loosely defined as 
\BEA
G&=&\textrm{number of isolated high-temperature connected regions}
\\
\nonumber
&-& \textrm{number of isolated low-temperature connected regions}. 
\label{eq:G}
\EEA
For instance, for a certain threshold, a hot spot will contribute $+1$
and a cold spot will contribute $-1$ to the genus. If a hot spot
contains a cold spot, the total contribution to the genus is zero.
The genus which is the global property of the
random field can be related to the integration of the local properties
of the field.
From the Gauss-Bonnet theorem, the genus of a closed curve $C$
being the boundary of a simply-connected region $\Omega_C$ 
which consists of $N$ arcs with exterior angles 
$\alpha_1,\alpha_2,...\alpha_N$ can be written in terms of
the geodesic curvature $\kappa_s$ and the Gaussian curvature $K$ as 
\BE
G=\f{1}{2 \pi}\Biggl[ \int_C \kappa_g ds+\sum_{i=1}^N \alpha_i+ \int_{\Omega_C}
K dA \Biggr ]. \label{eq:GB}
\EE
For a random field on the 2-dimensional Euclidean space $E^2$
where the N arcs are all geodesic segments (straight line segments),
$K$ and $\kappa_g$ vanish. Therefore, the genus is written as 
\BE
G_{E^2}=\f{1}{2 \pi}\sum_{i=1}^N \alpha_i \Bigr. \label{eq:F}
\EE
\\
The above formula is applicable to the locally flat spaces
such as $E^1\times S^1$ and $T^2$ which have $E^2$ as the 
universal covering space since 
$K$ and $\kappa_g$ also vanish in these spaces. In these 
multiply-connected spaces, the naive 
definition Eq.(\ref{eq:G}) is not correct for
excursion regions surrounded by a loop which cannot be contracted 
to a point. 
\\
\indent 
In order to compute the genus for a random field on a sphere $S^2$ 
with radius equal to 1, it is convenient to use 
a map $\psi$:$S^2-\{p_1\}-\{p_2\}\!\rightarrow\! S^1\times (0,\pi)$
defined as 
\BE
\psi:(\sin\theta \cos\phi,\sin\theta \sin\phi,\cos\theta)
\rightarrow (\phi,\theta),~~0\le\phi<2\pi,0<\theta<\pi,
\EE
where $p_1$ and $p_2$ denote the north pole and the south pole, respectively. 
Because $S^1\times (0,\pi)$ can be considered as locally flat spaces
($\phi,\theta$) with metric $ds^2=d\theta^2+d\phi^2$ which have 
boundaries $\theta\!=\!0,\pi$,
the genus for excursion regions that 
do not contain the poles surrounded by straight segments
in the locally flat ($\phi,\theta$) space is given
by Eq.(\ref{eq:F}). It should be noted that the straight segments do not 
necessarily correspond to the geodesic segments in $S^2$. 
If a pole is inside an excursion region 
and the pole temperature is above the threshold then the genus is 
increased by one. If the pole temperature is below the threshold, it
does not need any correction. Thus the genus for the excursions is 
\BE
G_{S^2}=\f{1}{2 \pi}\sum_i \alpha_i+N_p, \label{eq:Gs}
\EE
where $\alpha_i$ is the exterior angles at the intersection of two 
straight segments in the ($\phi,\theta$) space and $N_p$ is the number
of poles above the threshold. 
\\
\indent
Now consider an isotropic and homogeneous 
Gaussian random temperature field on a sphere $S^2$
with radius 1. Let $(x,y)$ be the local Cartesian coordinates on $S^2$ and 
let the temperature correlation function be
$C(r)\!=\!<\!(\Delta T/T)_0(\Delta T/T)_r\!>\!$ with $r\!=\!x^2+y^2$
and $C_0\!=\!C(0)\equiv \sigma^2$, where $\sigma$ is the standard
deviation and $C_2\!=\!-(d^2C/dr^2)_{r=0}$. Then the expectation
value of the genus for a threshold $\Delta T /T=\nu\sigma$ is given 
as \cite{Adler}
\BE
<G_{S^2}>=\sqrt{\f{2}{\pi}}\f{C_2}{C_0}\nu\textrm{e}^{-\nu^2/2}+\textrm{erfc}
\Biggl(\f{\nu}{\sqrt{2}}\Biggr ), \label{eq:aveGs}
\EE 
where erfc($x$) is the complementary error function. The first term in 
Eq.(\ref{eq:aveGs}) is equal to the averaged contribution for the excursions
which do not contain the poles while the second term in Eq.(\ref{eq:aveGs})
is the expectation value of $N_p$.  
\\
\indent
The mean contour length per unit area for an isotropic homogeneous 
Gaussian random field is \cite{Gott,Adler}
\BE
<s>=\f{1}{2}\Biggl ( \f{C_2}{C_0}\Biggr)^{\f{1}{2}}\textrm{e}^{-\nu^2/2},
\EE
and the mean fractional area of excursion regions for the field is
the cumulative probability of a threshold level,
\BE
<a>=\f{1}{2}\textrm{erfc}\Biggl(\f{\nu}{\sqrt{2}}\Biggr),
\EE
which gives the second term in Eq.(\ref{eq:aveGs}).
\\
\indent
\begin{figure}[tpb]
\centerline{\psfig{figure=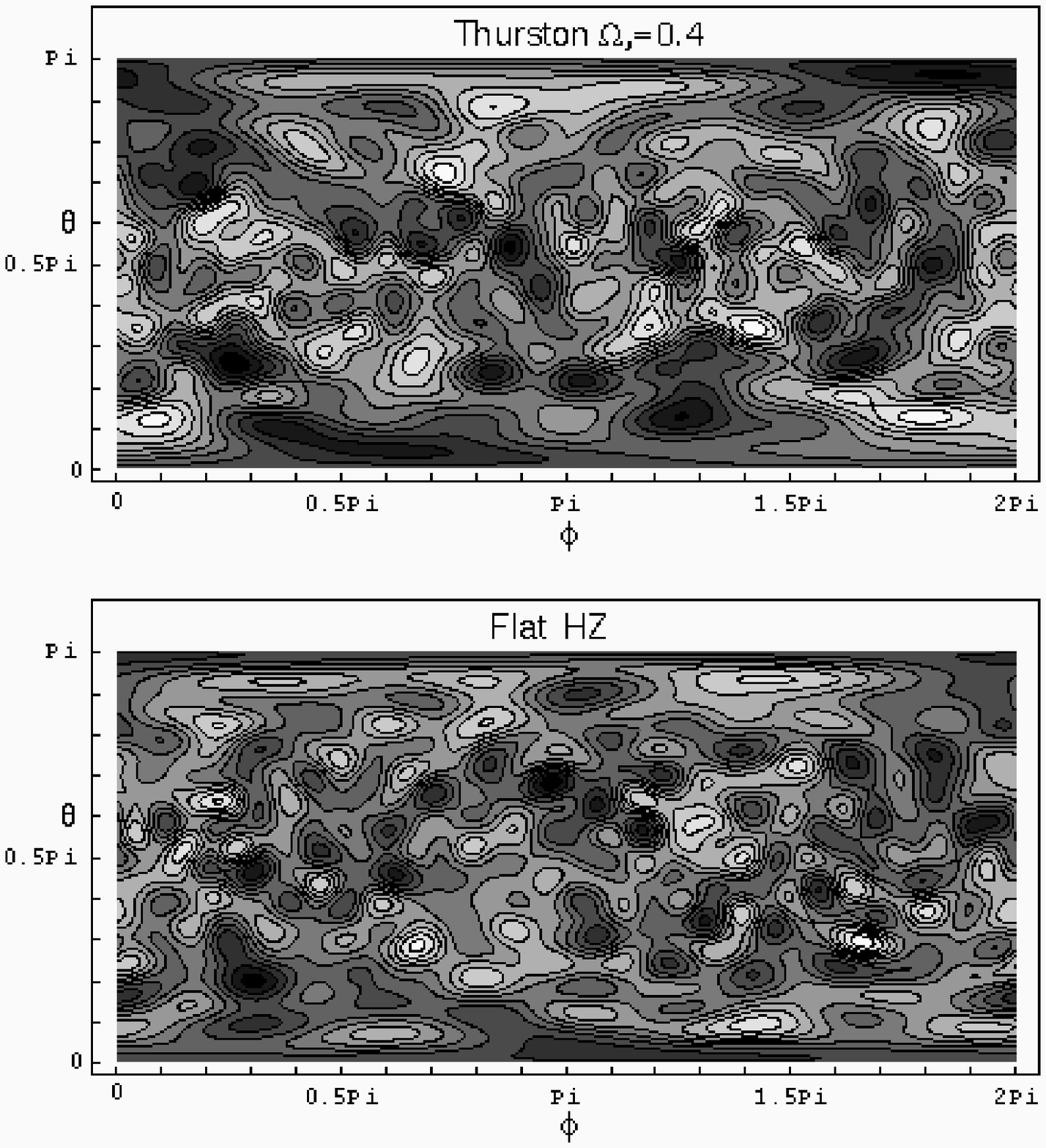,width=18cm}}
\caption{Contour maps of the 
CMB (not smoothed by the DMR beam) for the Thurston model 
$\Omega_0\!=\!0.4$ and 
a flat (Einstein-de-Sitter) 
Harrison-Zel'dovich model $C_l \propto
1/(l(l+1))$ in which all multipoles $l\!>\!20$
are removed.}

\label{eq:Sky}
\end{figure}
As in Sec.~III, the CMB anisotropy maps for the two CH
adiabatic models are produced by using eigenmodes $k<13$ and angular 
components $2\!\leq\!l\!\leq\!20$ for $\Omega_0\!=\!0.2$ and $0.4$. The
contribution of higher modes are approximately 7 percent and 10
percent for $\Omega_0\!=\!0.2$ and $0.4$, respectively. The initial
power spectrum is assumed to be the extended Harrison-Zel'dovich 
spectrum. The beam-smoothing effect is not included. 
For comparison, sky maps for the Einstein-de-Sitter model with
the Harrison-Zel'dovich spectrum $C_l\propto1/(l(l+1))$ are also
simulated. 
\\
\indent
In order to compute the genus and the contour length for each model,
10000 CMB sky maps on a 400$\times$200 grid in the $(\phi,\theta)$
space are produced. 
The contours are approximated by oriented straight
segments. The genus comes from the sum of the exterior angles at the
vertices of the contours and the number of poles at which the
temperature is above the threshold.
The total contour length is approximated by 
the sum of all the straight segments.  
Typical realizations of the sky map are shown in Fig.13. 
\\
\indent
Fig.14 and Fig.15 clearly show that the mean genuses and the mean total
contours for the two CH models 
are well approximated by
the theoretical values for the Gaussian models. This is a natural
result since the distribution of the expansion coefficients $b_{lm}$ 
is very similar to
the Gaussian distribution in the modest range.
On the other hand, at high and low threshold levels,
the variances of the total contour lengths
and the genuses are larger than that for the Gaussian models
that can be attributed to the nature of the distribution
function of $b_{lm}$. 
One can easily notice the non-Gaussian signatures from Fig.16 and Fig.17.
The excess variances for 
the Weeks model $\Omega_0\!=\!0.4$ compared with the Gaussian flat
Harrison-Zel'dovich model are observed at the absolute 
threshold level approximately $|\nu|\!>\!1.4$ for genus and
$|\nu|\!>\!0.6$ for total contour length.
If one assumes that the initial fluctuations are 
given by $(\Phi_{\nu}(0))^{-2}\!\propto\!\nu(\nu^2+1)$, the temperature
fluctuations for CH models can be described as Gaussian pseudo-random fields.
One can see from Fig.18 that the behavior of the 
variances of genus and total contour length for the Gaussian CH models
is very similar to that for the flat Harrison-Zel'dovich model 
and the variances at high and low threshold levels are considerably 
smaller than that for 
the non-Gaussian models. 
\\
\indent
Because the mean behavior for the two non-Gaussian CH models is well described
by the Gaussian models, the COBE DMR data 
which excludes grossly non-Gaussian models \cite{Colley,Kogut} 
cannot constrain the two CH models by the topological measurements.
However, one should take account of a fact that the signals in the
$10^o$ smoothed COBE DMR 4-year sky maps are comparable to the noises
\cite{Bennett} that makes it hard to detect the non-Gaussian signals
in the background fluctuations.  In fact, some recent works 
using different statistical tools have shown that the COBE DMR 4-year
sky maps are non-Gaussian \cite{Ferreira,Novikov,Pando} although 
some authors cast 
doubts upon the cosmological
origin of the observed non-Gaussian signals \cite{Bromley,Banday}. 
Thus the 
evidence of Gaussianity in the CMB
fluctuations is still not conclusive.
\begin{figure}
\centerline{\psfig{figure=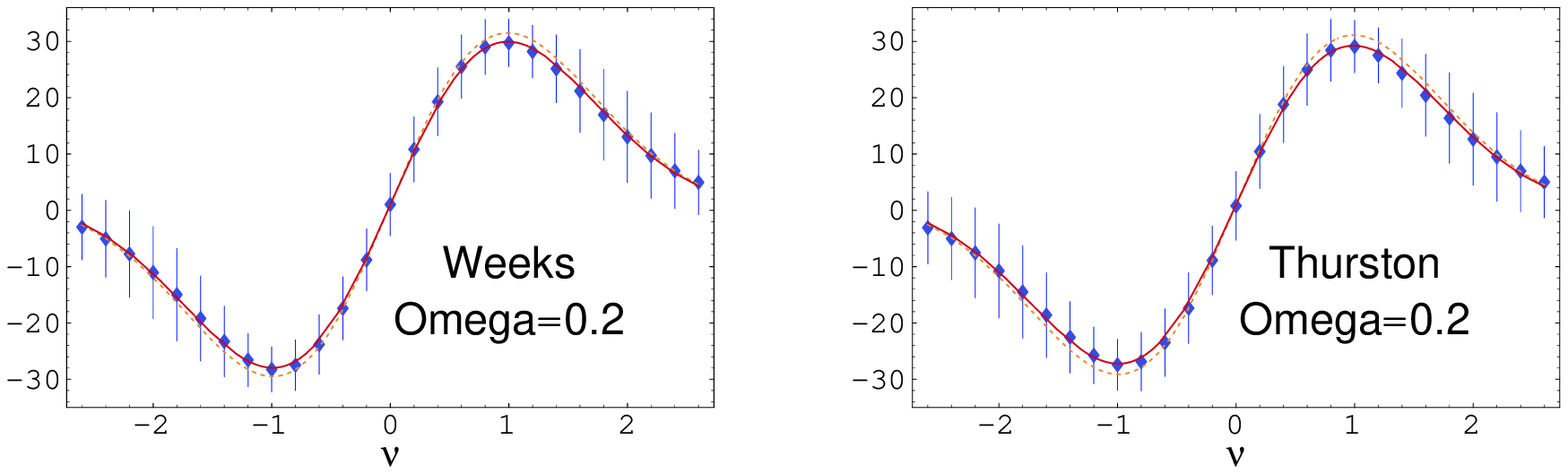,width=18cm}}
\caption{The mean genuses averaged over 100 realizations of the
initial fluctuations and 100 realizations of the base points
and $\pm 1 \sigma$ run-to-run variations at 27 threshold levels 
for the Weeks and the Thurston models with $\Omega_0\!=\!0.2$.
The dashed curves denote the mean values for a Gaussian model 
where $C_0$ and $C_2$ are obtained by assuming that the expansion
coefficients of the eigenmodes are random Gaussian numbers (the mean
is zero and the variance is proportional to $\nu^{-2}$). The solid 
curves denote the mean values for a Gaussian model that are 
best-fitted to that for CH models.}  
\label{eq:Genus0.2}
\end{figure}

\begin{figure}
\centerline{\psfig{figure=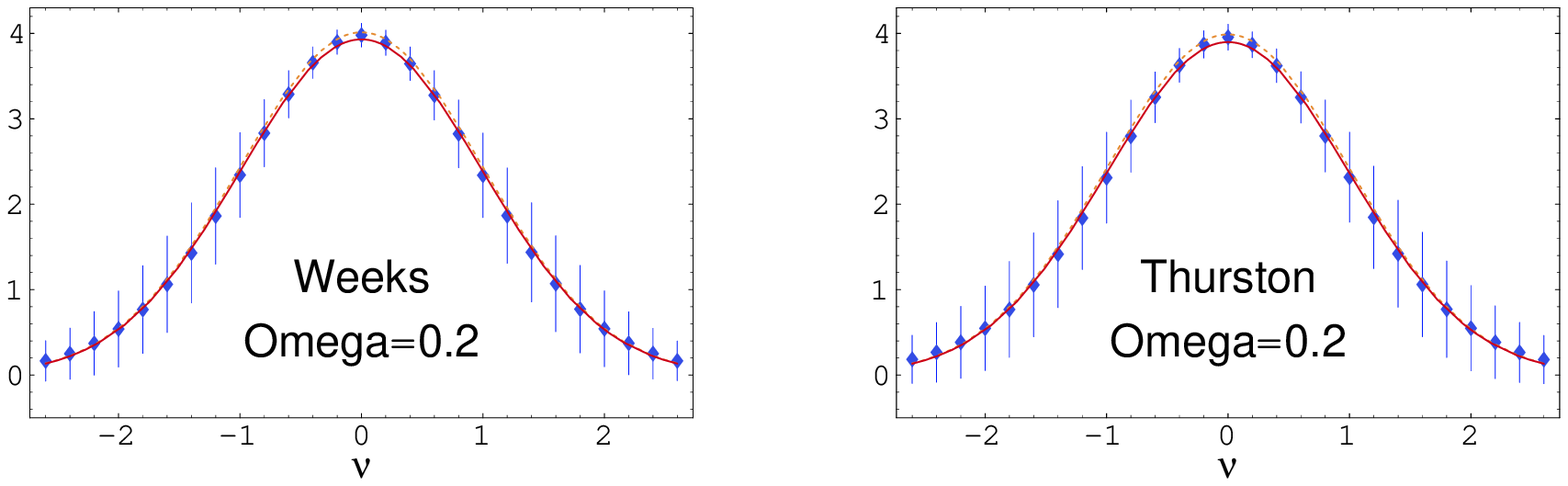,width=18cm}}
\caption{The mean contour lengths averaged over 100 realizations of the
initial fluctuations and 100 realizations of the base points
and $\pm 1 \sigma$ run-to-run variations at 27 threshold levels 
for the Weeks and the Thurston models with $\Omega_0\!=\!0.2$.
The dashed curves denote the mean values for a Gaussian model 
where $C_0$ and $C_2$ are obtained by assuming that the expansion
coefficients of the eigenmodes are random Gaussian numbers (the
mean is zero and the variance is proportional to $\nu^{-2})$. The solid 
curves denote the mean values for a Gaussian model that are 
best-fitted to that for CH models.}  
\label{eq:Length0.2}
\end{figure}

\begin{figure}
\centerline{\psfig{figure=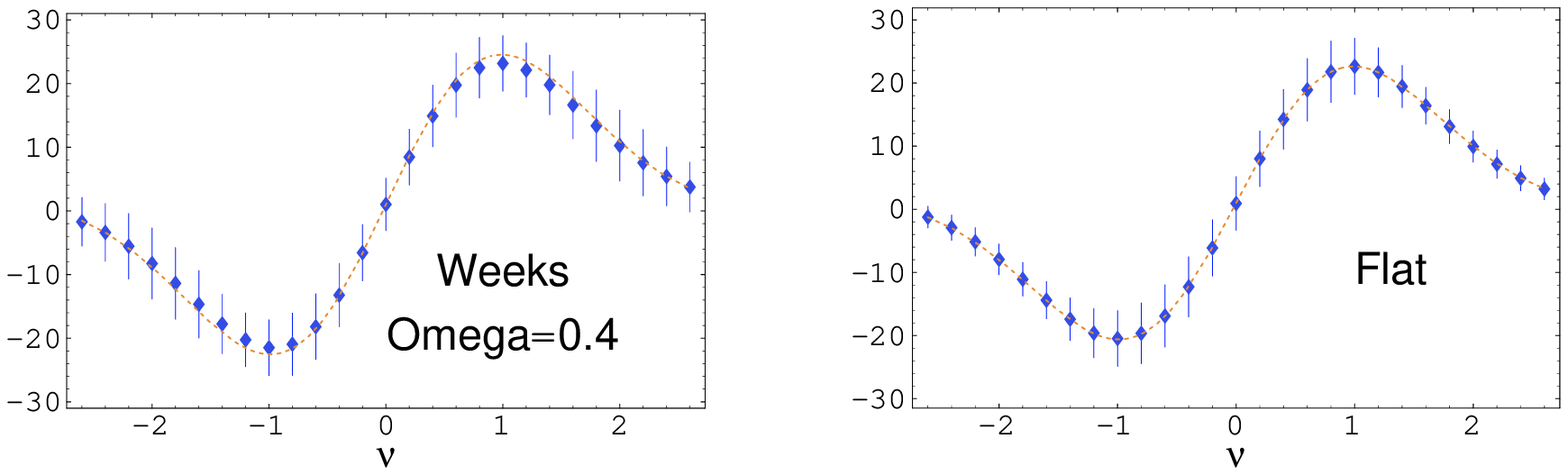,width=18cm}}
\caption{The mean genuses and $\pm 1 \sigma$ run-to-run variations at 27
threshold levels for a Weeks model with $\Omega_0\!=\!0.4$
averaged over 100 realizations of the
initial fluctuations and 100 realizations of the base points
and that for a flat Harisson-Zel'dovich model averaged over
10000 realizations.
The dashed curves denote the mean genuses for the corresponding 
Gaussian models.}
\label{eq:Genus0.4WandF}
\end{figure}

\begin{figure}[tpb]
\centerline{\psfig{figure=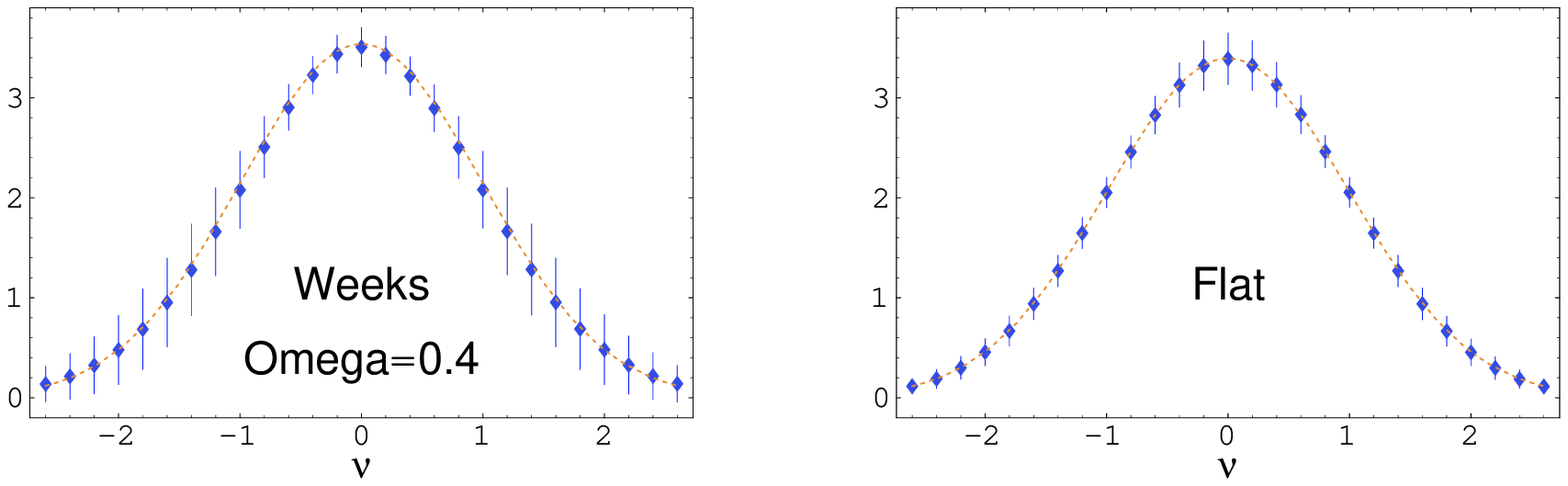,width=18cm}}
\caption{The mean total contour lengths  and $\pm 1 \sigma$ 
run-to-run variations at 27
threshold levels for a Weeks model 
with $\Omega_0\!=\!0.4$ averaged over 100 realizations of the
initial fluctuations and 100 realizations of the base points
and that for a flat Harisson-Zel'dovich model averaged over
10000 realizations.
The dashed curves denote the mean total contour lengths for the corresponding 
Gaussian models.}
\label{eq:Length0.4WandF}
\end{figure}

\begin{figure}[tpb]
\centerline{\psfig{figure=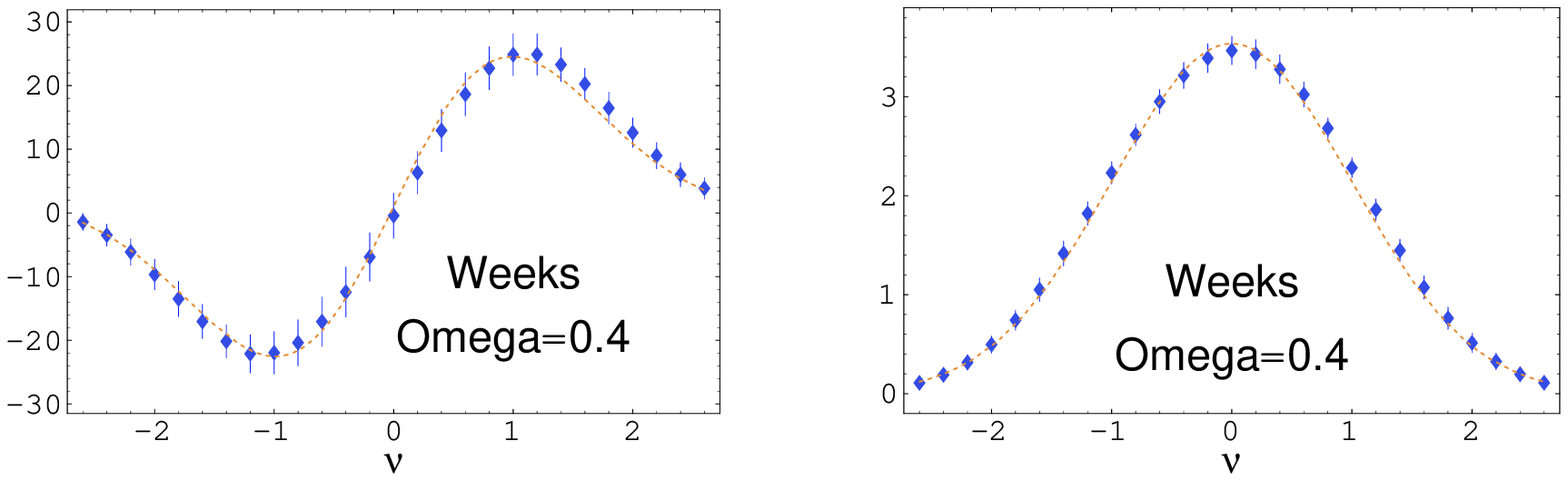,width=18cm}}
\caption{The mean  total contour lengths and genuses and $\pm 1 \sigma$ run-to-run variations at 27
threshold levels for a Weeks model 
with $\Omega_0\!=\!0.4$ averaged over 300 realizations of the base points.
Here it is assumed that the initial fluctuations deterministically 
satisfy  $(\Phi_{\nu}(0))^{-2}\!\propto\!\nu(\nu^2+1)$ so 
that the fluctuations 
are described by the Gaussian statistics.  
The dashed curves denote the mean total contour lengths and the mean
genuses for the corresponding Gaussian models.}
\label{eq:Phiconst0.4W}
\end{figure}
\pagebreak
\section{CONCLUSION}
In this paper, Gaussianity of the eigenmodes and 
non-Gaussianity in the CMB
temperature fluctuations in
two smallest CH(Weeks and Thurston) models are investigated. 
As shown in Sec.~II, it is numerically shown that the 
expansion coefficients of the two CH 
spaces behave as if they are random Gaussian numbers at
almost all the places. If one recognizes the Laplacian as the Hamiltonian 
of a free particle, each eigenmode is interpreted as a wavefunction
in a stationary state. The observed behavior is consistent with a
prediction of RMT which has been considered to be a good empirical
theory that describe the statistical properties of quantum mechanical 
systems whose classical counterparts are strongly
chaotic.
However, as we have seen, the global symmetries in the system
can veil the generic properties. 
For instance, some eigenmodes on the Thurston manifold 
have a $Z2$ symmetry at a point where the injectivity radius is
maximal. For these eigenmodes, the expansion coefficients are
strongly correlated;hence they can no longer considered to be random
Gaussian numbers. 
\\
\indent
Because the eigenmodes actually satisfy the periodic boundary conditions,
there are points on a sphere $S^2$ which are identified
with different points on $S^2$. These points form pairs
of circles which are identified by the periodic boundary conditions 
\cite{Cornish4}. 
If one could identify all the circles on a sphere, one would be able
to construct the corresponding CH space \cite{Weeks}. 
Similarly, if one could identify all the fixed points and the
corresponding symmetries, one would be able to construct a CH space
which have these symmetries.
The observed ``randomness'' in the eigenmodes is actually 
determined by these simple structures.
\\
\indent
In order to understand the symmetric structures of the CH spaces,
it is useful to choose an observing point (base point) at which
one enjoys symmetries as many as possible. However, in reality, 
there is no natural reason to consider fluctuations at only  
these particular points since the CH spaces are 
globally inhomogeneous. 
\\
\indent
Since the CMB fluctuations can be written in terms of a linear combination
of eigenmodes, the fluctuations in CH models are almost spatially 
``isotropic'' if averaged all over the
space except for very
limited places at which the eigenmodes have certain symmetries
provided that the eigenmodes are Gaussian.
The spatial ``isotropy'' implies that 
the contribution of non-diagonal terms in the two-point 
correlation functions are negligible.  Thus the 
validity of the statistical tests using
the angular power spectrum $C_l$ \cite{Aurich3,Inoue2,Inoue3,Cornish1} 
cannot be questioned
on the ground that the background space is anisotropic at a certain
point.
\\
\indent
If one assumes that the initial fluctuations are Gaussian
as in the standard inflationary scenarios, the temperature
fluctuations are described by isotropic non-Gaussian random fields 
since they are written in terms of a sum of products of two independent 
random Gaussian variables, namely the initial perturbations and the
expansion coefficients of the eigenmodes. 
The distribution functions of the expansion
coefficients $b_{lm}$ for the sky maps at large values 
are slowly converged to zero than the
Gaussian distribution with the same variance and the cosmic
variances are found to be larger than that of the Gaussian models.
\\
\indent
The increase in the variances are much conspicuous for topological
quantities at large or small threshold levels. On the other hand,
the mean behavior is well approximated by the Gaussian predictions.
Therefore, the obtained results agree with the COBE DMR 4-year maps 
analyzed in\cite{Colley,Kogut}. 
In real observations one has to 
tackle with what obscure
the real signals such as pixel noises, galactic contaminations, 
beam-smoothing effect and systematic calibration errors which have
not been considered in this paper.
The absence of large deviations from the mean 
values at large or small threshold levels in the current data
may be due to these effects, which will be much explored 
in the future work. 
\\
\indent
Although the recent observations seem to prefer the flat FRW models
with the cosmological constant, the evidence is not perfectly conclusive.
If one includes the cosmological constant for a fixed curvature radius,
the radius of the last scattering surface (horizon) at present in 
unit of curvature radius becomes large. Therefore the observable
imprints of the non-trivial topology of the
background space become much prominent.  
For instance, the number $N_f$ of copies of the fundamental domains inside
the last scattering at the present slice is approximately 27.9 
for a Weeks model with $\Omega_\Lambda\!=\!0.6$ and $\Omega_m\!=0.2\!$ 
whereas $N_f\!=\!4.3$ if $\Omega_\Lambda\!=\!0$ and $\Omega_m\!=0.8\!$. 
Thus we have still great possibilities in detecting the non-trivial
topology by the future satellite missions such as MAP and PLANCK which 
will provide us much better information on the statistical properties of
the real signals.  The large deviations of the topological
quantities from the mean values would be the good signals that 
indicate the hyperbolicity (negative curvature) and the 
finiteness (smallness) of the universe in addition to the direct 
observation of the periodic structures peculiar to each non-trivial 
topology (see \cite{Luminet} for recent developments). 
\\
\indent
\vspace{3cm}
\section*{Acknowledgments}
I would like to thank Jeff Weeks, Makoto Sakuma, Michihiko Fujii, and 
Craig Hodgson for answering many questions about symmetric
structures of compact hyperbolic 3-spaces and topology of 3-manifolds.
I would also like to thank N.J. Cornish, Naoshi Sugiyama and Kenji
Tomita for their informative comments.
The numerical computation in this work was carried out  
at the Data Processing Center in Kyoto University and 
Yukawa Institute Computer Facility. 
K.T. Inoue is supported by JSPS Research Fellowships 
for Young Scientists, and this work is supported partially by 
Grant-in-Aid for Scientific Research Fund (No.9809834). 

\end{document}